\documentclass[times,twocolumn,final]{article}

\usepackage{float}
\usepackage{framed,multirow}
\usepackage{amsmath}
\usepackage{amssymb}
\usepackage{latexsym}
\usepackage{graphicx}
\usepackage{url}
\usepackage{xcolor}
\usepackage{tikz}
\usepackage{hyperref}
\usepackage{subcaption}
\definecolor{newcolor}{rgb}{.8,.349,.1}

\usepackage{booktabs}
\usepackage{diagbox}
\usepackage{siunitx}
\usepackage{colortbl}
\usepackage[a4paper, margin=0.5in]{geometry}

\usepackage[numbers]{natbib}

\title{``\emph{No negatives needed}'': weakly-supervised regression for interpretable tumor detection in whole-slide histopathology images}

\author{
Marina D'Amato$^1$, Jeroen van der Laak$^1$, Francesco Ciompi$^1$ \\
\small $^1$Computational Pathology Group, Radboud University Medical Center, Nijmegen, The Netherlands \\
\small Corresponding author: \texttt{marina.damato@radboudumc.nl}
}
\date{}

\begin{document}
\twocolumn[
\begin{@twocolumnfalse}
\maketitle
\begin{abstract}
Accurate tumor detection in digital pathology whole-slide images (WSIs) is crucial for cancer diagnosis and treatment planning. Multiple Instance Learning (MIL) has emerged as a widely used approach for weakly-supervised tumor detection with large-scale data without the need for manual annotations. However, traditional MIL methods often depend on classification tasks that require tumor-free cases as negative examples, which are challenging to obtain in real-world clinical workflows, especially for surgical resection specimens.
We address this limitation by reformulating tumor detection as a \emph{regression} task, estimating tumor percentages from WSIs, a clinically available target across multiple cancer types.
In this paper, we provide an analysis of the proposed weakly-supervised regression framework by applying it to multiple organs, specimen types and clinical scenarios.
We characterize the robustness of our framework to tumor percentage as a noisy regression target, and introduce a novel concept of \emph{``amplification technique''} to improve tumor detection sensitivity when learning from small tumor regions.
Finally, we provide interpretable insights into the model's predictions by analyzing visual attention and logit maps. 
\end{abstract}
\vspace{1cm}
\end{@twocolumnfalse}
]

\section{Introduction}
\label{sec1}

\begin{figure*}[t]
    \centering
    \includegraphics[width=1.0 \linewidth]{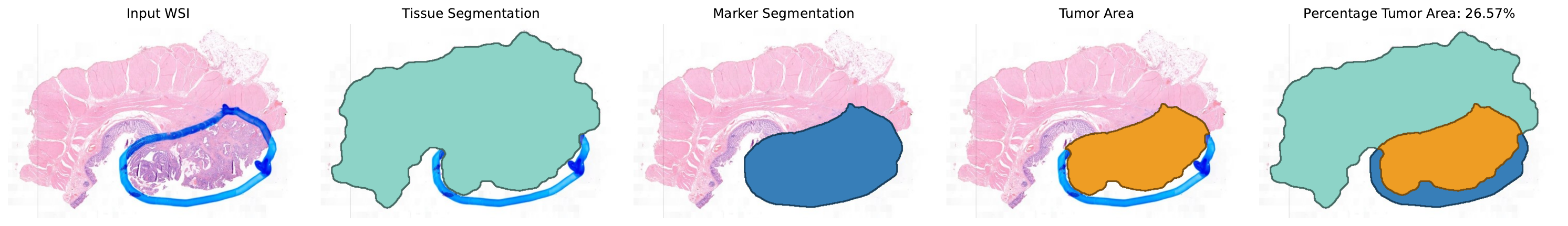} 
  \caption{Visual example of a procedure to extract an approximate tumor percentage area from a slide used in molecular diagnostics procedures. From a  coarse annotation of the tumor area, often provided with a pen marker, we can derive the tumor percentage via simple image analysis steps: 1) segmentation of foreground tissue versus background, 2) identification of the pen marker and area filling, 3) intersection of tissue and marker area, 4) calculation of the tumor percentage.}
  \label{fig:example_molecularlab}
\end{figure*}

Early and accurate cancer diagnosis is crucial for successful treatment and improved patient outcomes. Traditionally, histopathological analysis, where pathologists examine tissue sections under a microscope, serves as the gold standard for diagnosing tumors. However, this process is labor-intensive, time-consuming and subject to inter-reader variability (\cite{Wang2020}, \cite{Krieger1994}, \cite{Marron-Esquivel2023}).

The digitization of whole-slide images (WSIs) has enabled the development of automated analysis methods using artificial intelligence (AI) and deep learning, which can enhance diagnostic efficiency and accuracy \citep{vanderLaak2021a}. Automating tumor detection can alleviate the workload on pathologists potentially improving the diagnostic process and reducing turnaround time. Traditional deep learning models for tumor detection rely heavily on fully-supervised approaches, necessitating detailed pixel- or patch-level annotations (\cite{xiaPatchlevelTumorClassification2018}, \cite{Halicek2019}, \cite{Cruz-Roa2014}). This dependency on extensive annotations poses challenges in scalability and generalizability, primarily due to the scarcity of annotated datasets and the limited availability of expert pathologists.

Weakly-supervised learning (WSL), which requires only coarse annotations or slide-level labels, has emerged as a viable alternative to fully-supervised models, substantially reducing the annotation burden. While various WSL approaches exist, including those based on image compression \citep{Tell21, Aswo21} and streaming techniques \citep{Pinc19, Doop23}, Multiple Instance Learning (MIL) (\cite{Carbonneau2018}, \cite{fatimaComprehensiveReviewMultiple2023}) has emerged as the predominant framework in digital pathology. MIL treats each WSI as a "bag" containing image patches ("instances"), where only the bag-level label is available (e.g., tumor presence or absence). MIL has been successfully applied to binary classification tasks across various cancer types, including prostate cancer and basal cell carcinoma (\cite{campanellaClinicalgradeComputationalPathology2019}), breast cancer (\cite{ilseAttentionbasedDeepMultiple2018}), colorectal cancer (\cite{chikontwe2020multiple}), and skin cancer (\cite{Geij24}), achieving performance comparable to fully-supervised models without the need for manual annotations. 

Despite its advantages, most investigations have remained confined to classification problems with categorical outcomes, such as tumor presence or absence. A significant limitation in tumor detection formulated as a weakly-supervised binary classification task is the need for large datasets containing both positive (tumor present) and negative (tumor absent) cases. However, obtaining WSIs completely tumor-free in real-world clinical settings can be challenging as most resected specimens typically contain at least some degree of tumor. While biopsies and lymph nodes include both tumor-positive and tumor-negative cases, focusing on these tissues alone introduces biases and limits model generalizability. Unlike resections, these samples do not capture the full morphological diversity of tumors or their surrounding microenvironments. Since surgical specimens offer a broader and more representative range of tumor variations, they are a more suitable focus for large-scale applications. However, the scarcity of completely negative resections hinders the scalability of classification-based tumor detection models. 

To address this challenge, researchers have explored alternative targets for weakly-supervised learning, such as tumor percentages. In clinical practice, pathologists routinely delineate the tumor bed on WSIs as a preparatory step for molecular diagnostics, as shown in the first step of Figure\ref{fig:example_molecularlab}. This involves marking the approximate tumor area, often using a pen marker, and visually estimating the percentage of tumor cells within the annotated region to ensure adequate tumor content for molecular testing. 

Building on this workflow, \cite{Lerousseau2021a} proposed a weakly-supervised learning approach that leverages the tumor cell percentage provided by pathologists as a proxy to train a weakly-supervised segmentation algorithm derived from MIL. Their method uses the percentage of tumor cells \emph{within} the tumor bed to assign proxy instance-level labels to image patches to approximate segmentation maps that match the global percentage estimate. However, the tumor cell percentage does not always accurately represent the true spatial extent of the tumor in the slide. Moreover, their work primarily focused on frozen sections and showed limited success when applied to data from routine diagnostics based on formalin-fixed, paraffin-embedded (FFPE) tissue from two test cohorts. 

We propose to shift the focus from \emph{tumor cell} percentages to \emph{tumor area} percentages, under the hypothesis that tumor area percentage provides a more informative learning signal for tumor detection, as it better reflects the actual distribution of the tumor within the tissue.
By leveraging the pathologist-annotated tumor areas as a starting point, we employ simple image analysis steps (illustrated in Figure \ref{fig:example_molecularlab}, more information in the Supplementary Material) to calculate the percentage of tumor area relative to the entire tissue section.
We then use this percentage as a direct target to train a regression MIL framework, an approach that has been previously used for stromal tumor-infiltrating lymphocytes (\cite{Schirris2021}), inferring gene expression profiles (\cite{Weitz2021}, \cite{Graziani2022a}), survival prediction (\cite{Chen2022}), and molecular biomarker prediction (\cite{elnahhasRegressionbasedDeepLearningPredicts2024}), but never for tumor detection.

\subsection{Our contributions}
Our key contributions are as follows. 
First, we introduce a novel application of Multiple Instance Learning (MIL) for tumor detection with weakly-supervised regression, circumventing the need for extensive annotations and tumor-free cases. We explore the applicability of this approach on various tissue types and compare various weakly-supervised methods in a regression setting, examining the impact of different models and pooling strategies. 

Second, we acknowledge that visual estimation of the tumor area made by pathologists is subject to variability and noise. To address this, we conduct a robustness analysis to evaluate the performance when building our model under different levels of \emph{synthetic noise} in the tumor percentage targets. This analysis is crucial for ensuring the reliability of our approach in real-world clinical settings, where such variability is unavoidable. 

Additionally, we introduce and analyze the effectiveness of a non-linear transformation of the target, which we refer to as the \emph{``amplification technique''} to enhance the training process of our regression models, particularly when dealing with cases involving small lesions. 

Lastly, we assess the interpretability of our models by comparing raw prediction maps based on instance-level predictions with attention maps derived from attention scores. These heatmaps provide valuable insights into the model's decision-making process, aligning with the clinical need for explainable AI in medical image analysis. We conduct a quantitative evaluation of the interpretability performance by comparing the heatmaps with ground truth annotations. 

\begin{figure}[t]
    \centering
    \includegraphics[width=1. \linewidth]{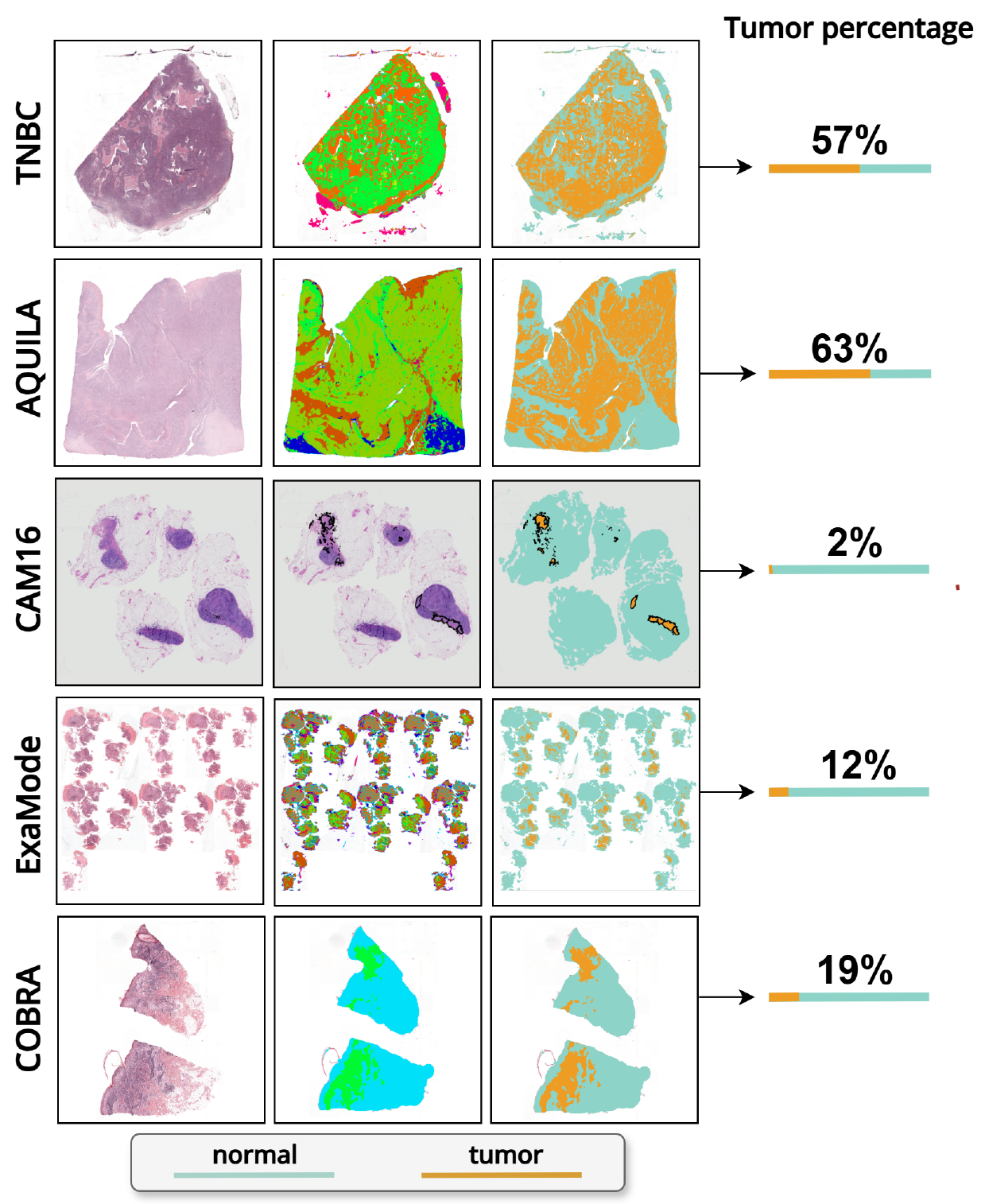} 
  \caption{Examples of WSIs paired with manual annotations (when available) or tumor segmentation masks, illustrating the computed tumor percentages. The first column presents the input WSIs, the second column displays either manual annotations (for CAM16) or the raw output of the segmentation algorithm, and the third column showcases the binarized maps highlighting the tumor regions.}
  \label{fig:pcts_examples}
\end{figure}

\section{Materials}
We evaluated weakly-supervised regression models on five datasets consisting of WSIs stained with hematoxylin and eosin (H\&E), covering a diverse range of specimen types, including surgical resections, lymph nodes, and biopsies from different organs. Figure \ref{fig:pcts_examples} provides a visual overview of the input data and computed tumor percentages.

\paragraph{Breast TNBC}
The TNBC dataset \citep{Balk20} includes 595 cases of triple negative breast cancer (TNBC) surgical resections from an equal amount of patients. This dataset was constructed from a multicenter, retrospective cohort study and encompasses patients diagnosed with TNBC between 2006 and 2014 from several hospitals in the Eastern Netherlands. The slides were scanned using a Pannoramic 1000 DX scanner (3DHISTECH) at a pixel resolution of 0.24 $\mu$m. Tumor percentages in this dataset were computed using the publicly available HookNet algorithm \citep{Rijt21}, as manual annotations were not available. HookNet demonstrated strong segmentation performance on invasive tumors, achieving Dice scores of 0.9 and 0.91 for IDC and ILC, respectively, making it well-suited for estimating tumor percentages in this dataset. All the slides in this dataset contain cancer with percentages ranging from 2\% to 66\%. The mean and median percentage of tumor across the slides are 25\% and 24\% respectively, indicating a relatively even distribution of tumor burden without significant skewness. 

\paragraph{CAMELYON16}
The CAMELYON16 dataset (\cite{EhteshamiBejnordi2017}) (CAM16) consists of 399 WSIs of lymph node sections (one slide per patient) from two medical centers in the Netherlands: Radboud University Medical Center (RUMC) and University Medical Center Utrecht (UMCU). RUMC images were scanned using a digital slide scanner (Pannoramic 250 Flash II; 3DHISTECH) with a \(20 \times\) objective lens, resulting in a specimen-level pixel size of \(0.243 \, \mu\text{m} \times 0.243 \, \mu\text{m}\). UMCU images were scanned with a NanoZoomer-XR Digital slide scanner C12000-01 (Hamamatsu Photonics) using a \(40 \times\) objective lens, producing a pixel size of \(0.226 \, \mu\text{m} \times 0.226 \, \mu\text{m}\). 
This dataset includes pixel-level annotations for both macro-metastases and micro-metastases which we used to compute tumor percentages. Of the 399 slides, 160 contain metastases while the remaining depict normal tissue. The percentages of tumor in tumor slides range from 0.003\% to 71\%, with 91 slides having a percentage less than 1\%. The mean percentage of tumor across the tumorous slides is approximately 5\% while the median is only 0.4\%, indicating a skewed distribution towards lower tumor percentages, highlighting the challenge of detecting small metastases in this dataset.

\paragraph{EXAMODE Colon}
The ExaMode colon dataset \citep{Cont24} consists of 8556 H\&E-stained colorectal biopsy WSIs cut from 6556 paraffin blocks of 3501 patients collected from the Radboud University Medical Center between 2000 and 2009. As the diagnoses were reported at the block level, slides from the same block were combined into a single new slide (``packed'' slide \citep{Aswo}) by minimizing the area of background between sections. Pathology reports associated with these slides provide labels for five classes: normal, hyperplastic polyps, low-grade dysplasia (lgd), high-grade dysplasia (hgd), and cancer. For this study, we grouped normal and hyperplastic polyps into a single ``normal'' class, resulting in two classes: normal (4546 cases) and abnormal (lgd, hgd, and cancer; 2010 cases). Tumor percentages in abnormal cases were derived from segmentation maps created using a colon tissue segmentation algorithm \citep{Bokh23} which demonstrated high performance, achieving a Dice score of 0.89 for tumor segmentation. The percentages in this dataset range from 0.033\% to 62\%, with a mean of 14\% and median of 10\%.
The small size and sparse distribution of tumor areas in these slides pose significant challenges for accurate tumor detection.

\begin{figure*}[ht!]
    \centering
    \includegraphics[width=0.9\textwidth]{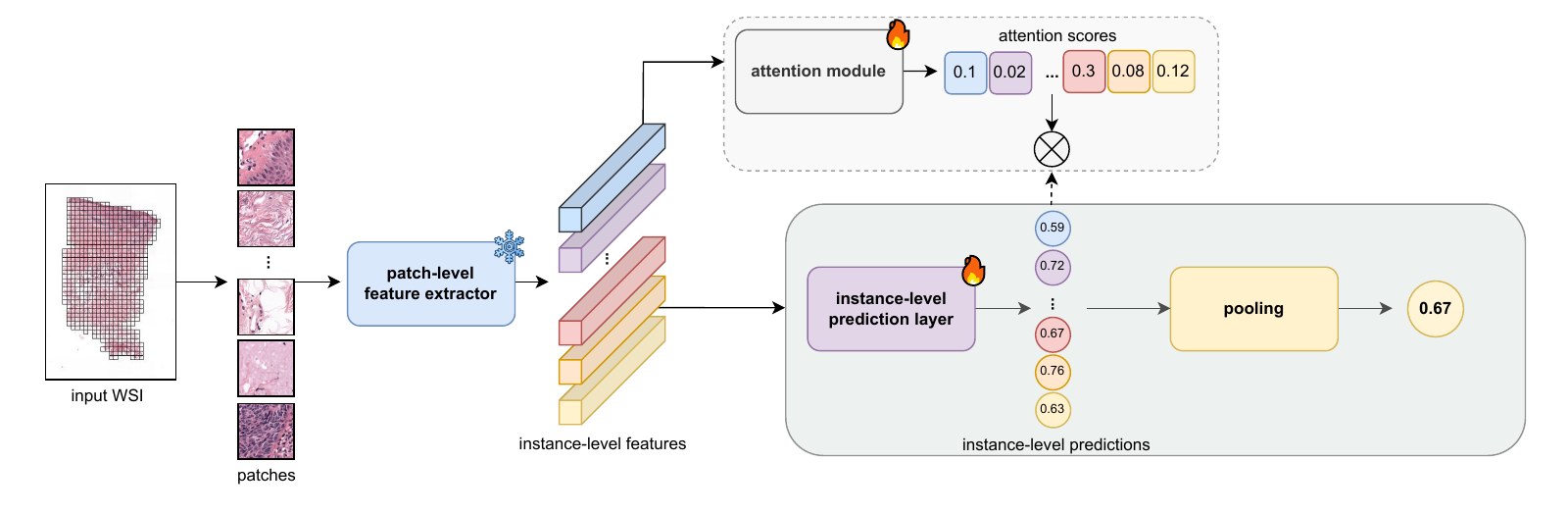} 
    \caption{Overview of the instance-based multiple instance learning (MIL) framework. The instance-based approach processes individual patches from the input image independently, making predictions at the instance level before aggregating them.}
    \label{fig:instance_embedding}
\end{figure*}

\paragraph{AQUILA Colon}
The AQUILA colon dataset includes 571 WSIs of colon tissue from surgical resections, with one slide per patient. These slides were scanned at two institutions: Radboud University Medical Center (RUMC) and Laboratorium Pathologie Oost-Nederland (LabPON). As for the EXAMODE cohort, tumor percentages were computed using the same colon tissue segmentation algorithm (\citep{Bokh23}), grouping the segmentation labels of high-grade dysplasia (HGD), low-grade dysplasia (LGD), and tumor into a single “abnormal” class, while all other tissue types were classified as “normal.” All the slides in this dataset contain cancer with percentages ranging from 0.078\% to 62\%, with a mean percentage of 13.33\% and a median of 12.27\%. 

\paragraph{COBRA}
The COBRA dataset (\cite{Geij24}) consists of 5147 slides from 4066 patients obtained from Radboud University Medical Center between 2016 and 2020. It includes patients diagnosed with Basal Cell Carcinoma (BCC), epidermal dysplasia (actinic keratosis or Bowen's disease), or benign conditions. Among the slides, 2661 contain BCC tumors, while 2476 are classified as non-BCC. All slides were scanned using a 3DHistech Pannoramic 1000 scanner at \(20 \times\)magnification (pixel resolution 0.24 $\mu$m). Tumor percentages were computed using an in-house skin tumor segmentation algorithm (more information in the Supplementary Material) and range from 0.01\% to 89\%, with a mean of 16\% and a median of 10\%. This distribution suggests a moderate range of tumor burden within the BCC-positive slides.

\section{Methods}
We investigated adaptations of the classical MIL classification approach \citep{campanellaClinicalgradeComputationalPathology2019} for regression tasks, using the tumor percentage within a WSI as the continuous target.
In this section, we first retrace the main concepts of MIL, we then introduce the proposed approaches for MIL regression, the analysis that we performed on the impact of noisy labels as well as the motivation and the effect of the ``amplification technique''.

\subsection{Multiple Instance Learning}
\label{methods}
In Multiple Instance Learning, a WSI is treated as a \textit{bag} containing multiple image patches, each referred to as an \textit{instance} $x_i$, where $i=0, 1, ..., N$ and $N$ represents the total number of patches in the slide. The task is to predict a bag-level target $y$ based on the instances, where each patch $x_i$ does not have a direct label but contributes to the overall prediction. Instead of assigning individual labels to each patch, the model learns a bag-level prediction based on the aggregated information from all patches within the WSI.
MIL methods are typically divided into two categories: \emph{instance-based} and \emph{embedding-based} approaches, depending on how the individual patches are processed and aggregated. The \emph{instance-based} method starts by extracting features from each image patch in the WSI. These features are then passed through an instance-level prediction layer, which generates a prediction for each patch independently. After obtaining the instance-level predictions, these are combined into a single prediction for the entire slide using a pooling mechanism, which can be as simple as mean pooling or more sophisticated, such as attention-weighted pooling. 

In this study, we focus on \emph{instance-based} approaches to MIL (see Figure \ref{fig:instance_embedding}) because it aligns with the framework used in \cite{Lerousseau2021a}, where proxy-labels are assigned to individual instances, and instance-level predictions are made. Additionally, we prioritize interpretability in our approach, which is facilitated by instance-based MIL since it allows for direct visualization of individual patch predictions, enabling a clearer understanding of how the model makes its decisions.

\subsection{Multiple instance learning to regress tumor percentage}
We evaluate four different MIL approaches, each adapted to the task of tumor percentage regression and to the instance-based formulation of MIL. These methods vary in their strategies for aggregating information from individual patches to predict the tumor percentage for the entire slide.

\paragraph{MeanPool}
The first method, \emph{Meanpool MIL} (MeanPool), is a classical approach that uses the average of all patch predictions within a slide as the final tumor percentage, providing a simple and intuitive way to combine information from all patches. 

\paragraph{Attention-based MIL}
The second method, \emph{Attention-based MIL} (ABMIL), introduced by \cite{ilseAttentionbasedDeepMultiple2018}, leverages the attention mechanism as a pooling operator. It assign weights (attention scores) to each patch, indicating its relative importance for the final prediction. Higher scores signify that the model considers the corresponding patch to be more informative about the tumor content. ABMIL uses these attention scores to create a weighted average of the patch predictions, focusing more on informative regions within the WSI. 

\paragraph{CLAM}
Building upon ABMIL's interpretability, the third method, \emph{Clustering Constrained Attention MIL (CLAM)}, incorporates an instance clustering loss function (\cite{Lu2020}). This loss function encourages the model to identify a subset of patches with high attention scores (likely containing tumor) and a separate subset with low attention scores (likely non-tumor). These patches are then used for additional training with pseudo-labels, potentially improving the model's ability to differentiate between tumor and non-tumor regions. In this study, we adapt the original embedding-based CLAM formulation to an instance-based approach. 

\paragraph{WeSEG}
\label{method_4}
Lastly, we examined \emph{Weakly Supervised Segmentation} (WeSEG) (\cite{Lerousseau2021a}). While previous methods employ a two-step approach, where a frozen encoder maps all patches to low-dimensional embeddings, WeSEG trains the encoder end-to-end to generate patch-level tumor probability maps using weak labels. It employs a recursive training mechanism to generate proxy labels and iteratively refines the model's predictions. At every step, it creates proxy labels by assigning label 1 to the $p$\% patches with the highest probability and label 0 to the remaining ones, where $p$ is the target tumor percentage. While WeSEG was not explicitly optimized to predict tumor percentages, we can derive percentage estimates from its patch-level predictions. Specifically, we convert patch-level probabilities to binary predictions using a threshold optimized per experiment, then calculate the tumor percentage as the ratio of patches classified as tumor to the total number of tissue patches. This post-processing step allows us to compare the predicted tumor percentage directly with the target annotation.

\subsection{Effect of noisy targets}
Pathologists' annotations of the tumor bed provide a valuable basis for estimating tumor area percentages. However, these annotations are inherently noisy due to their reliance on coarse manual delineations and visual estimation. In this work, we benefit from more precise tumor area estimates based on detailed manual annotations or AI-assisted delineation techniques (\cite{Rijt21}, \cite{Bokh23}). To account for the inherent variability in clinical practice, we introduce synthetic noise into the tumor percentage labels during training. This reflects the imprecision that naturally arises from visual estimation, allowing us to assess the robustness of the framework to noisy inputs, while testing and validation are performed using the true, unaltered labels.

We simulated three levels of noise, ranging from mild to substantial deviations, to model potential underestimation or overestimation by pathologists in real-world clinical scenarios. For each training example, we added noise to the true percentages by sampling from uniform distributions: [-0.1, 0.1] for mild noise ($\pm$10\%), [-0.3, 0.3] for moderate noise ($\pm$30\%), and [-0.5, 0.5] for substantial noise ($\pm$50\%). 

\begin{figure}[ht]
    \centering
    \includegraphics[width=1. \linewidth]{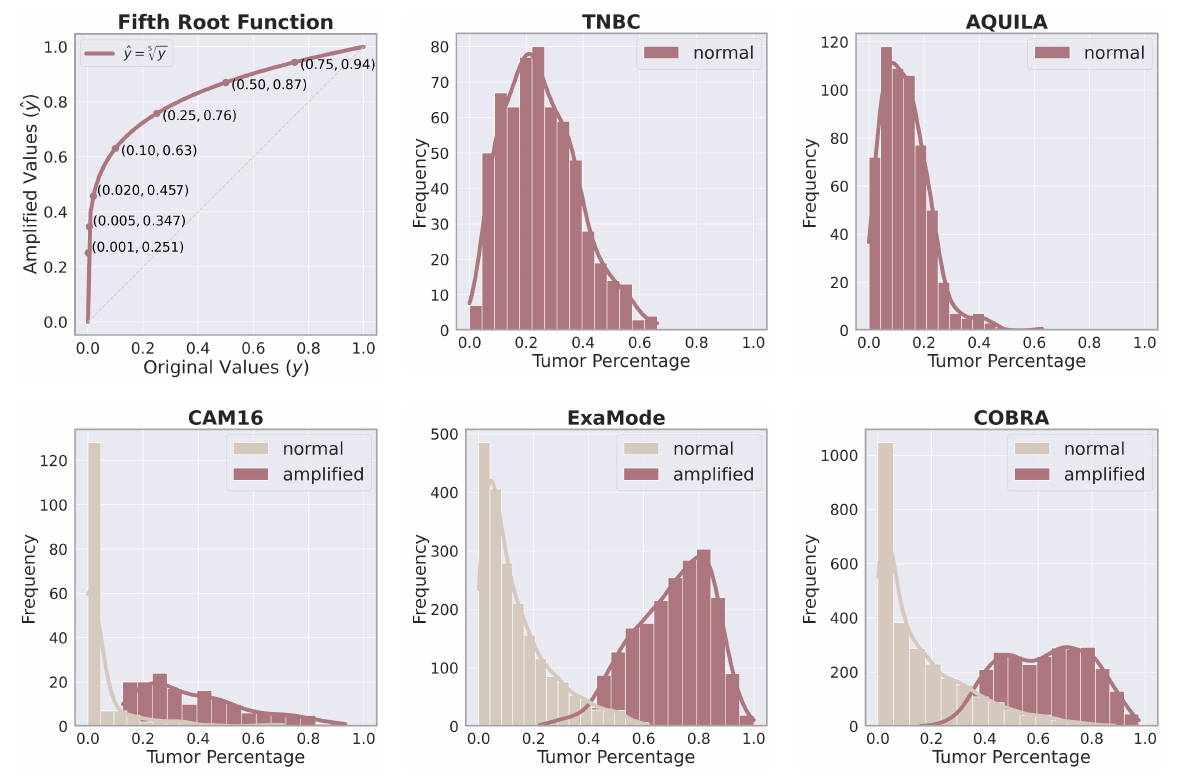} 
  \caption{The first panel illustrates the fifth root transformation, with annotated points demonstrating the effect of amplification on selected values, particularly enhancing lower percentages. The other five panels display the distribution of tumor percentages across various cohorts, shown only for tumorous slides (excluding negative cases), before and after the \emph{amplification technique}.}
  \label{fig:pcts_distribution}
\end{figure}

\subsection{Target amplification}
In the context of our regression task for tumor detection, cases with small lesions or low tumor percentages can be particularly difficult for the model to distinguish from tumor-free cases, potentially leading to reduced performance. To address this challenge, we employ a \emph{target amplification} technique designed to enhance the model's sensitivity to small tumor percentages.

This technique involves transforming the target labels using a root transformation $\hat{y} = \sqrt[n]{y}$ that expands lower-end values (small tumor percentages), making subtle differences between normal tissue and areas with small tumor lesions more discernible. For this work, we selected the fifth root transformation ($n=5$), based on empirical results demonstrating that it strikes a good balance between amplifying small tumor values without overly distorting higher tumor percentages. 

This transformation is particularly valuable for datasets such as CAM16, COBRA, and ExaMode consisting of biopsies and lymph nodes, which usually include either small tumor percentages or entirely tumor-free regions. Figure~\ref{fig:pcts_distribution} illustrates the distributions of tumor percentages before and after amplification, demonstrating how the transformation expands the range of small values, enhancing the separation between tumor-free and low-tumor cases. On the other hand, datasets such as AQUILA and TNBC consist of surgical resections with no tumor-free cases, where tumor areas are relatively large. In these datasets, amplification is unnecessary and was not applied in our analysis, as the clear distinction between tumor regions does not require further enhancement.

\section{Experiments}
\paragraph{Data preprocessing}
We first segmented the tissue from the background using a pre-trained tissue segmentation network (\cite{Bandi2019a}). From the segmented tissue regions, we extracted non-overlapping 256x256 patches at a spatial resolution of 0.5 $\mu$m using the \textit{hs2p} library (\cite{Grisi}). 
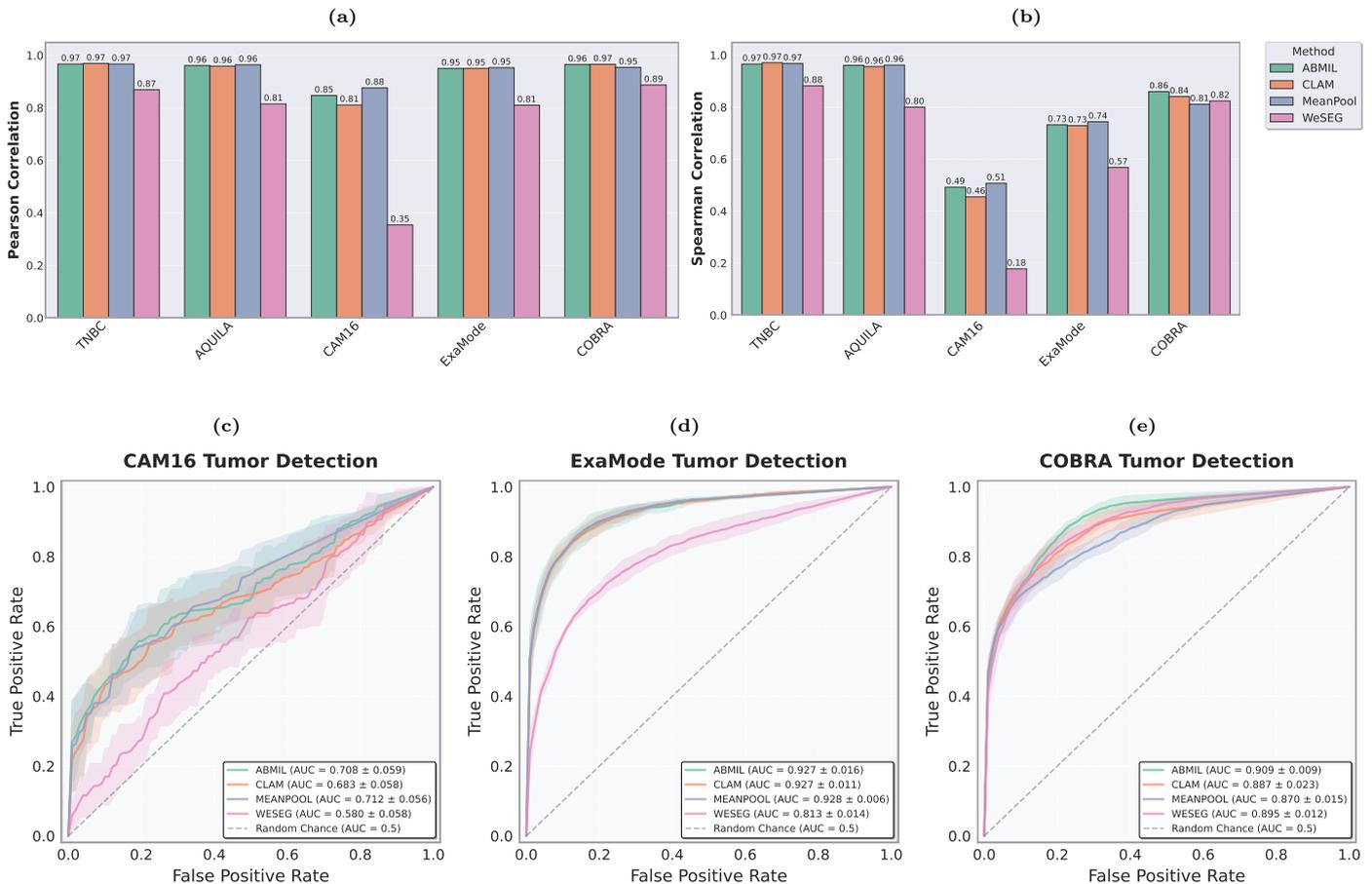
\begin{figure*}[h!]
    \centering

    \begin{minipage}{0.5\textwidth}
        \centering
        \begin{tikzpicture}
            \node[anchor=south west, inner sep=0] (image) at (0,0) {\includegraphics[width=\textwidth]{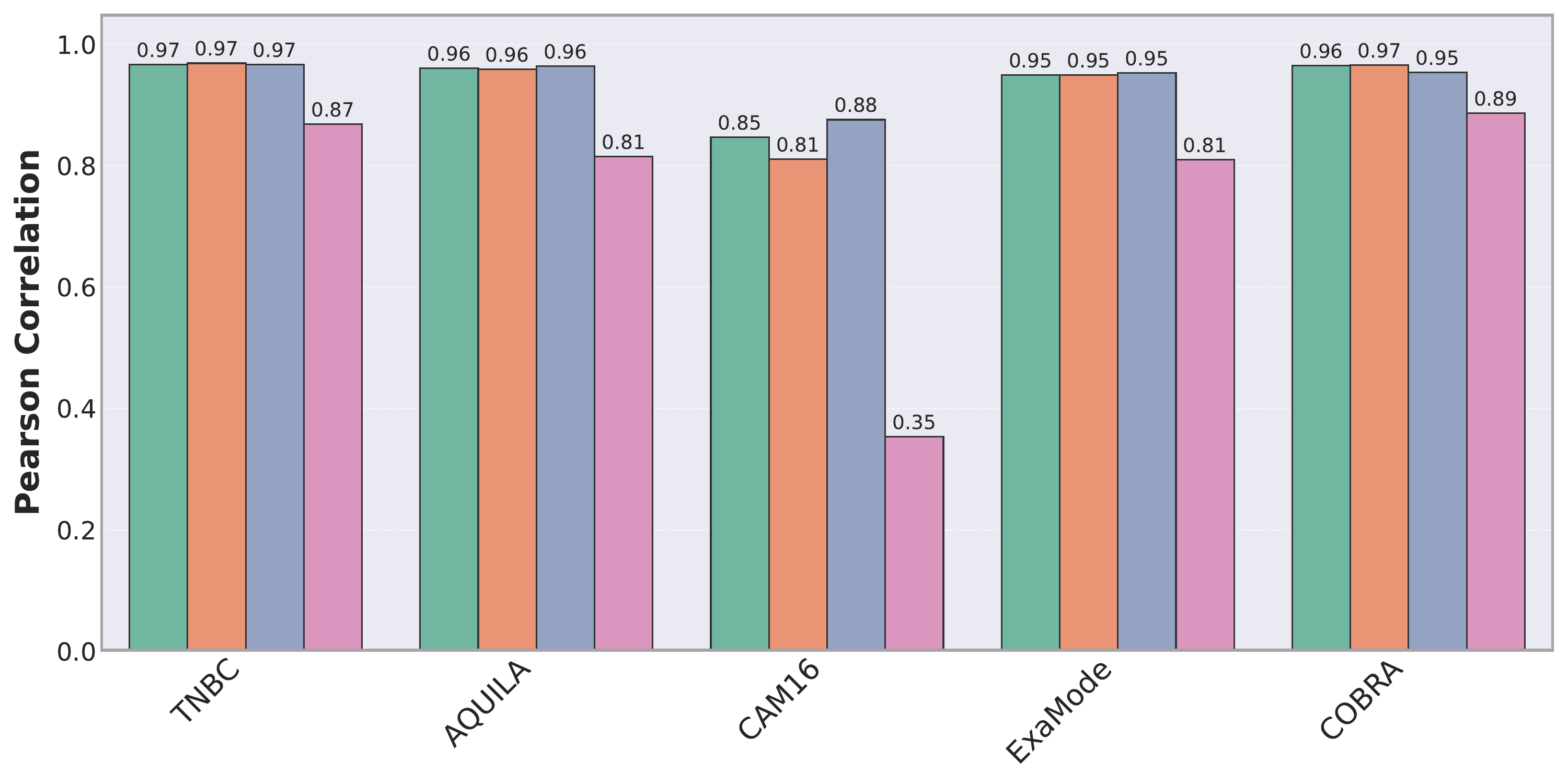}};
            \node[anchor=south, xshift=0.5\textwidth, yshift=0em] at (image.north west) {\scriptsize\textbf{(a)}}; 
        \end{tikzpicture}
        \label{fig:bar1}
    \end{minipage}%
    \hfill
    \begin{minipage}{0.5\textwidth}
        \centering
        \begin{tikzpicture}
            \node[anchor=south west, inner sep=0] (image) at (0,0) {\includegraphics[width=\textwidth]{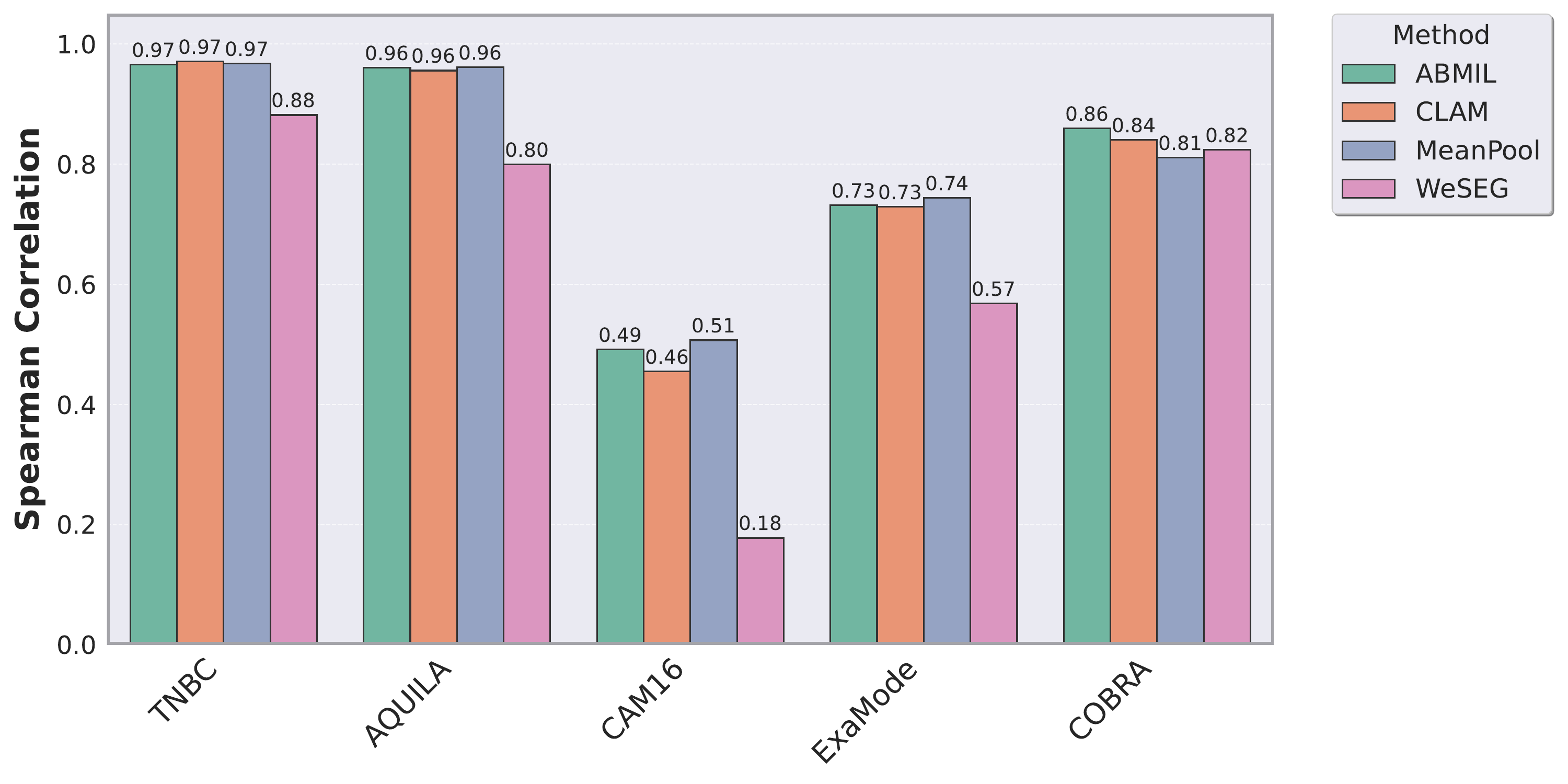}};
            \node[anchor=south, xshift=0.5\textwidth, yshift=0em] at (image.north west) {\scriptsize\textbf{(b)}}; 
        \end{tikzpicture}
        \label{fig:bar2}
    \end{minipage}
    \begin{minipage}{0.33\textwidth}
        \centering
        \begin{tikzpicture}
            \node[anchor=south west, inner sep=0] (image) at (0,0) {\includegraphics[width=\textwidth]{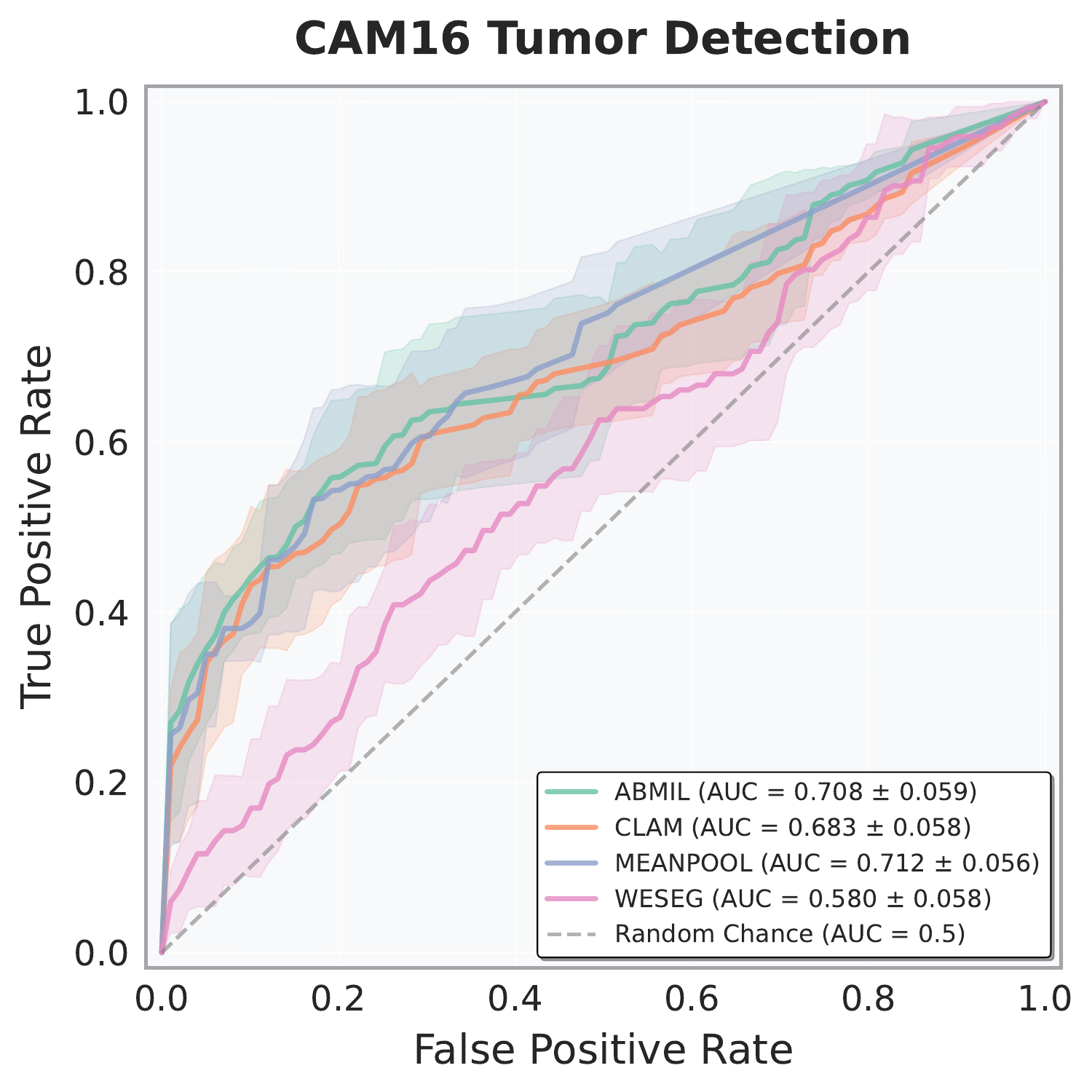}};
            \node[anchor=south, xshift=0.5\textwidth, yshift=0em] at (image.north west) {\scriptsize\textbf{(c)}}; 
        \end{tikzpicture}
        \label{fig:roc1}
    \end{minipage}%
    \hfill
    \begin{minipage}{0.33\textwidth}
        \centering
        \begin{tikzpicture}
            \node[anchor=south west, inner sep=0] (image) at (0,0) {\includegraphics[width=\textwidth]{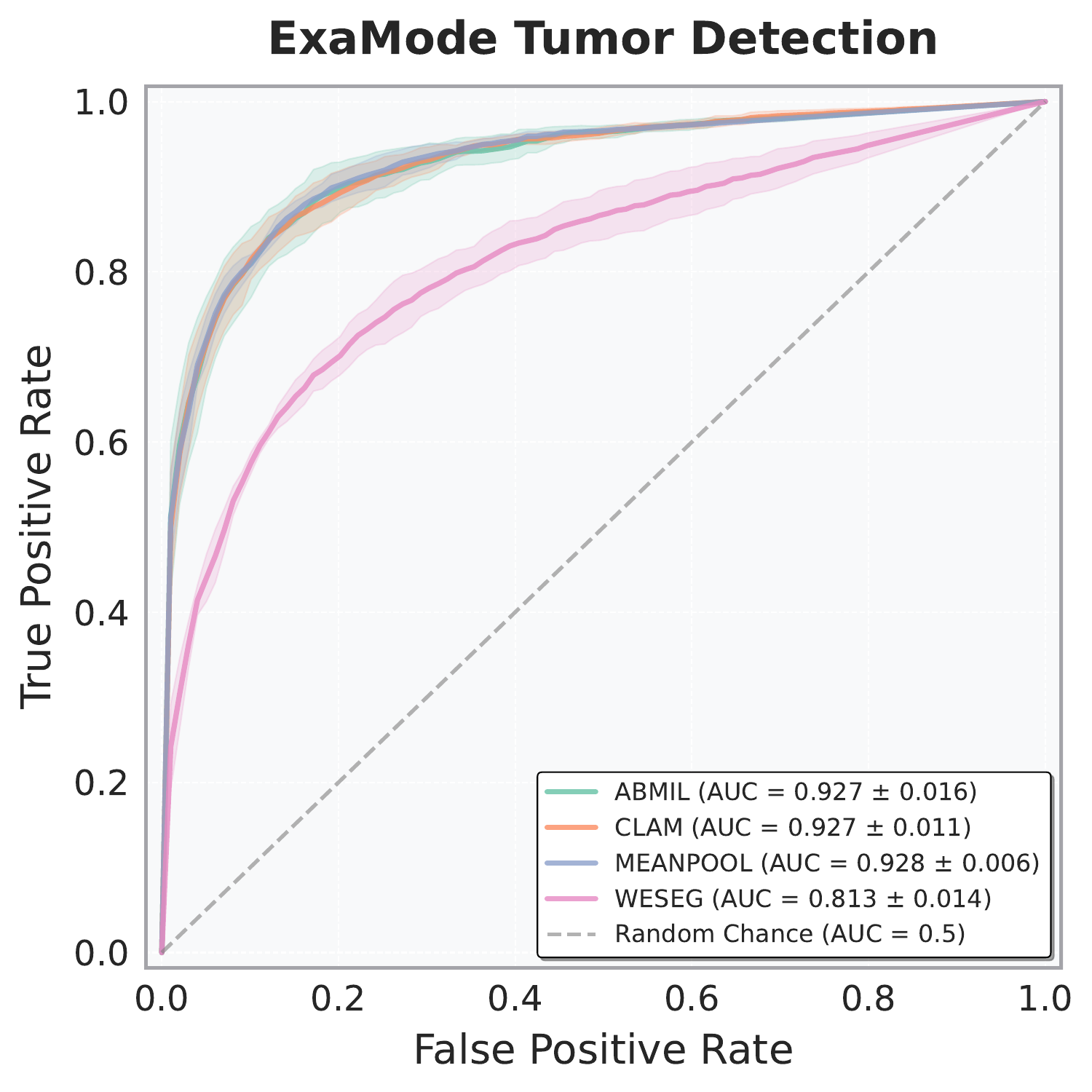}};
            \node[anchor=south, xshift=0.5\textwidth, yshift=0em] at (image.north west) {\scriptsize\textbf{(d)}}; 
        \end{tikzpicture}
        \label{fig:roc2}
    \end{minipage}%
    \hfill
    \begin{minipage}{0.33\textwidth}
        \centering
        \begin{tikzpicture}
            \node[anchor=south west, inner sep=0] (image) at (0,0) {\includegraphics[width=\textwidth]{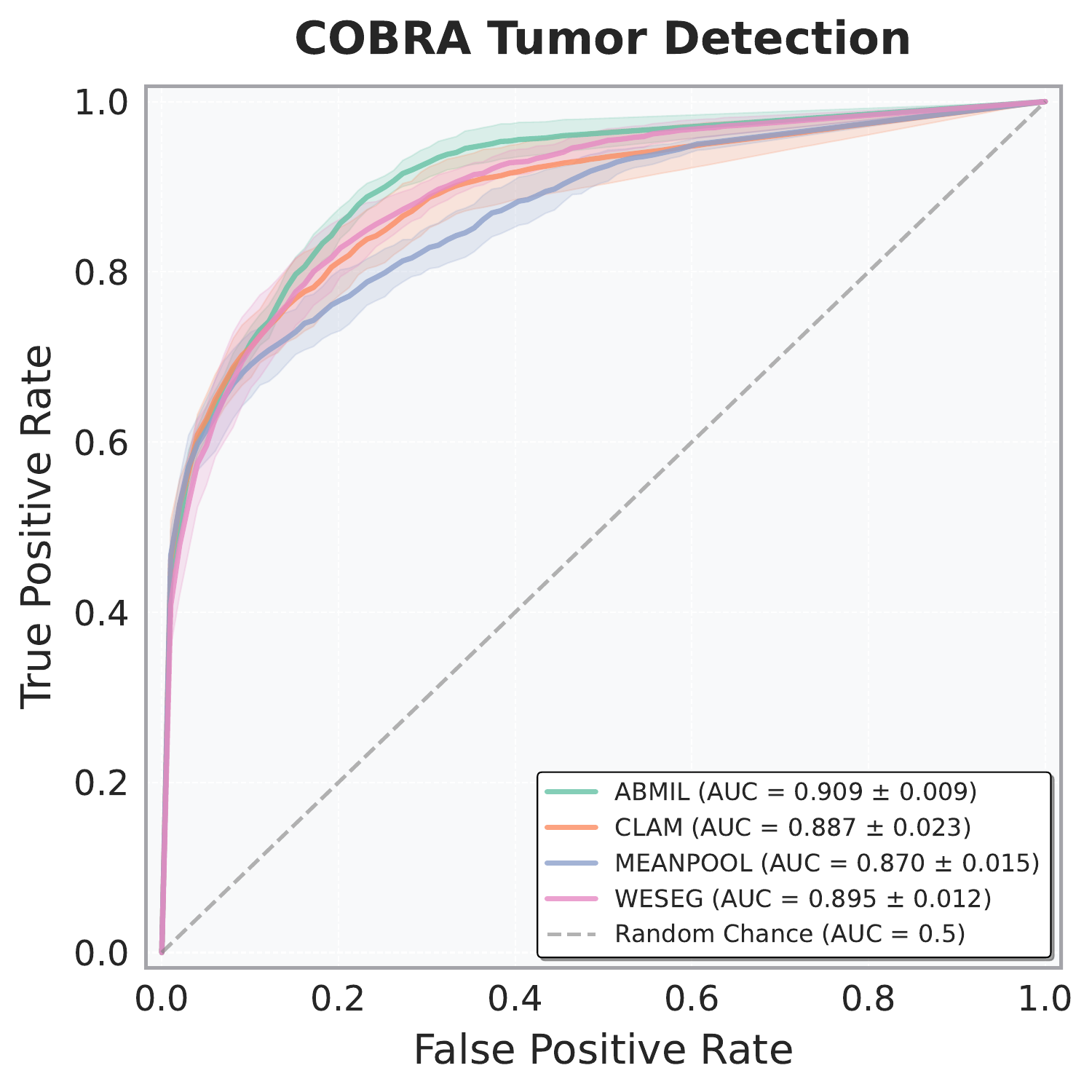}};
            \node[anchor=south, xshift=0.5\textwidth, yshift=0em] at (image.north west) {\scriptsize\textbf{(e)}}; 
        \end{tikzpicture}
        \label{fig:roc3}
    \end{minipage}

    \caption{Comparison of bar plots and ROC curves across methods and datasets. (a) Pearson correlation coefficients for different methods across the five datasets, indicating the strength of linear relationships between predicted and true tumor percentages. (b) Spearman correlation coefficients across the five datasets, showcasing the rank-based correlation between predictions and ground truth. (c-e) Receiver Operating Characteristic (ROC) curves for tumor detection with AUC scores for CAM16 (c), ExaMode (d), and COBRA (e), illustrating the trade-off between true positive and false positive rates. The shaded areas represent the confidence intervals over the 5-fold cross-validation, and the mean AUC values with standard deviations are reported for each dataset. These results are based on models trained using true tumor percentages.}
    \label{fig:combined_plots_normal}
\end{figure*}
\paragraph{Model Training}
With the exception of WeSEG, all investigated methods follow a two-step process involving a frozen convolutional neural network (CNN) encoder and a final classification layer (see Figure \ref{fig:instance_embedding}). In the first step, we used a modified ResNet50 pretrained on ImageNet as a patch-level feature extractor, following the approach described by \cite{Lu2020}, where the model was adapted by applying adaptive mean-spatial pooling after the 3rd residual block. The encoder’s weights were frozen, meaning they were not updated during training. The second step involved training the final layer for tumor percentage estimation.
The models were trained by minimizing the mean squared error (MSE) loss between the target and predicted tumor percentages with the Adam optimizer. Training included L2 weight decay set at $10^{-5}$ and a learning rate of $2\times10^{-4}$ was used. Training proceeded for at least 50 epochs and up to a maximum of 200 epochs, with early stopping implemented if the validation loss plateaued.

Unlike previous methods that rely on fixed feature extraction, WeSEG operates directly on the image patches using a model that learns to extract features during training. Here, the pre-trained ResNet50 architecture was used for end-to-end training, where the feature representations were continuously refined with each iteration. At each training step, 30 tiles were randomly sampled from each whole-slide image, as done in \cite{Lerousseau2021a}. Due to computational limitations, we restricted the sampling to a single whole-slide image per step. Based on the original work, data augmentation techniques were applied which included random vertical and horizontal flips, and color jittering with defined parameters for brightness, contrast, saturation, and hue (as implemented in \cite{Lerousseau2021a}). The model adopted the Adam optimizer with binary cross-entropy loss, with weight decay of $10^{-5}$ and a learning rate of 0.0005. Training spanned a maximum of 100 epochs, incorporating early stopping with a 50-epoch patience.

All experiments were conducted using a 5-fold cross validation strategy for each dataset to assess model performance generalizability. For each fold, we first selected the test set, which consisted of 20\% of the total slides. This ensured that, across all 5 folds, the test sets collectively represented the entire dataset, with no overlap between test sets. The remaining 80\% of the slides were then split into training (85\%) and validation sets (15\%). Stratified sampling was employed during the partitioning process to ensure that the distribution of tumor percentages remained consistent across the training, validation, and test sets. For cases containing multiple slides, where a "case" refers to all the slides from a single patient, all slides belonging to the same case were assigned to the same fold. Additionally, for CAM16 and COBRA, we performed separate training and validation on the official sets to obtain results comparable to those reported in literature.

\paragraph{Evaluation metrics}
We evaluated the regression task using the Spearman's and Pearson's correlation coefficients to quantify how closely our models' predictions align with the reference standard. Additionally, for datasets containing negative cases, we used receiver operating characteristic (ROC) curve analysis with the area under the curve (AUC) to assess tumor detection performance. In this analysis, binary labels (tumor vs. no tumor) were derived from tumor percentages to classify slides as positive or negative. 
Additionally, following \cite{Lerousseau2021a}, we used AUC to evaluate the interpretability of attention scores and instance-level predictions by comparing them against pathologists' ground-truth annotations or tumor segmentation masks. This comparison helps determine if attention heatmaps or instance-level predictions offer advantages for interpretability in regression tasks.

\section{Results}
\begin{figure*}[]
    \centering
    \includegraphics[width=1.0 \linewidth]{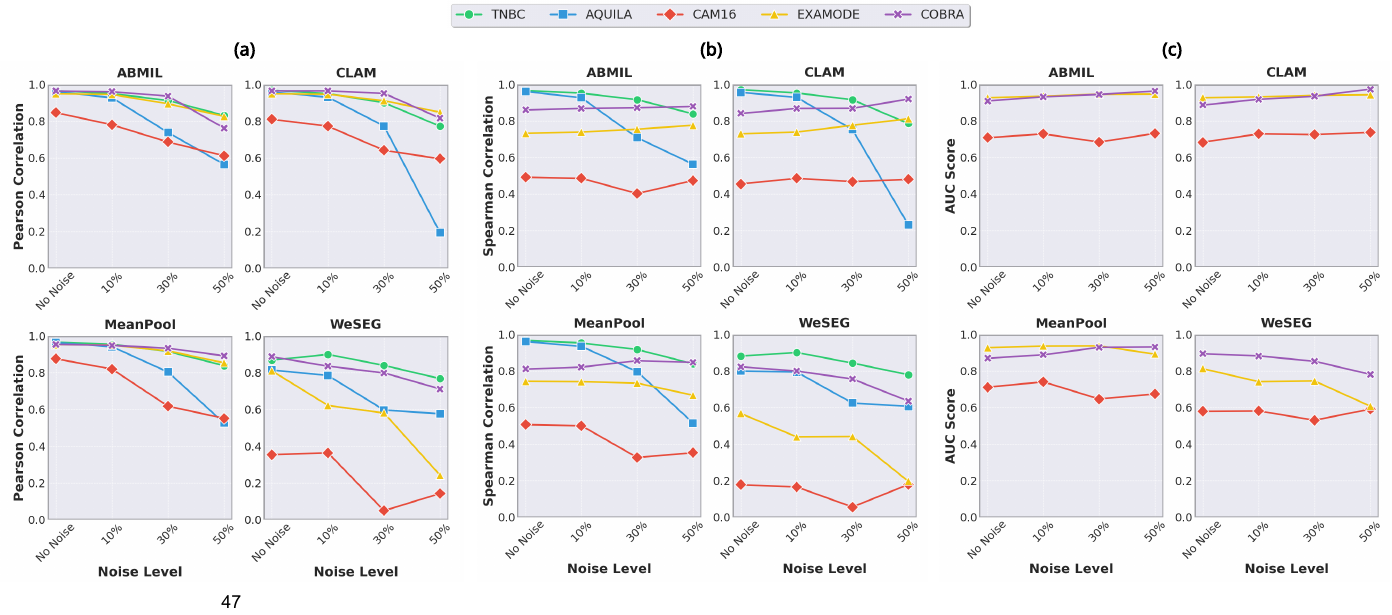} 
  \caption{Variation of Pearson, Spearman, and AUC metrics across different noise levels. Subplots (a), (b), and (c) depict line plots showing how the Pearson correlation, Spearman correlation, and AUC, respectively, change as noise levels increase. These metrics highlight the robustness of different methods to increasing levels of synthetic noise.}
  \label{fig:noise_comparison}
\end{figure*}

\subsection{Assessing tumor percentage prediction}
\label{subsec1}
Figure \ref{fig:combined_plots_normal}(a-b) present the Pearson and Spearman correlation results for various methods across multiple datasets. Overall, most methods demonstrate very strong correlations, highlighting their effectiveness in aligning predicted tumor percentages with the ground truth. However, WeSEG shows notably poorer performance compared to the other methods. This difference arises because WeSEG was not explicitly optimized to predict tumor percentages; rather, tumor percentages were derived from patch-level predictions using a binarization step, as described in section \ref{method_4}. This additional processing step introduces potential inaccuracies, as errors in the segmentation or imprecisions when converting to a percentage can impact the final result.

In the AQUILA and TNBC datasets, all methods (except WeSEG) achieve high Pearson ($r > 0.95$) and Spearman ($\rho > 0.95$) correlations, demonstrating robust performance in both value prediction and rank ordering. Surprisingly, MeanPool performs comparably to more complex methods such as ABMIL and CLAM, suggesting that methods with simpler pooling mechanisms can be effective in contexts where tumor areas are relatively large, such as in surgical resections.

For the ExaMode dataset, while Pearson correlations remain strong ($r > 0.95$), the Spearman correlations are lower ($\rho \approx 0.73$). This discrepancy highlights potential issues such as the presence of outliers which may cause the Pearson correlation to appear overly optimistic. Specifically, a detailed analysis reveals a significant number of tumor-free cases with predicted tumor percentages greater than zero (see the scatter plots in the Supplementary Material). For COBRA, all methods show strong performance, with Pearson correlations ranging from 0.95 to 0.97 for the attention-based methods. However, the Spearman correlations are slightly lower ($\rho \approx 0.85$), suggesting that the models may not consistently maintain the correct rank order of tumor percentages despite their strong linear relationships.

The CAM16 dataset poses unique challenges due to its heavily skewed tumor percentage distribution, with numerous samples containing extremely low tumor burdens (percentages $< 1\%$). This makes it difficult for models to discern small differences in tumor percentage values, which is reflected in lower Spearman correlations ($\rho < 0.51$) despite strong Pearson correlations ($r > 0.81$). WeSEG particularly underperforms on this dataset due to its reliance on post-processed patch-level predictions, which are less effective in detecting extremely small tumor regions.

\subsection{Assessing tumor detection capabilities}
\label{subsec2}
In this subsection, we assess the tumor detection capabilities on CAM16, ExaMode, and COBRA datasets, which contain both tumor and normal cases. The ROC curves and corresponding AUC scores for each method are presented in Figure \ref{fig:combined_plots_normal}(c-d-e).

On the CAM16 dataset, ABMIL achieves the strongest performance with an AUC of 0.716, followed by MeanPool (0.707) and CLAM (0.698). WeSEG showed notably lower performance with an AUC of 0.588. These results, however, are substantially lower compared to the state-of-the-art performance reported for binary classification approaches on this dataset. This performance gap highlights the increased complexity of tumor percentage regression when compared to traditional binary classification tasks, where learning to distinguish between tumor and normal cases is more straightforward.

Most methods show substantially higher performance on the ExaMode dataset. ABMIL, CLAM, and MeanPool achieve nearly identical performance with AUC scores ranging from 0.926 to 0.928. WeSEG, while showing improved absolute performance compared to CAM16, still underperforms with an AUC of 0.807. This suggests challenges in extracting a reliable tumor detection signal from weakly supervised segmentation maps, limiting the performance and usability of the method. 

On the COBRA dataset, all methods achieve strong performance. ABMIL leads with an AUC of 0.908. Interestingly, WeSEG showed competitive performance here with an AUC of 0.889, followed by CLAM (0.880) and MeanPool (0.870). While these results are promising, they remain below the reported benchmarks for binary tumor detection tasks on this dataset using classification labels, consistent with our observations reported on CAM16.

\begin{figure}[t]
    \centering
    \includegraphics[width=1.0 \linewidth]{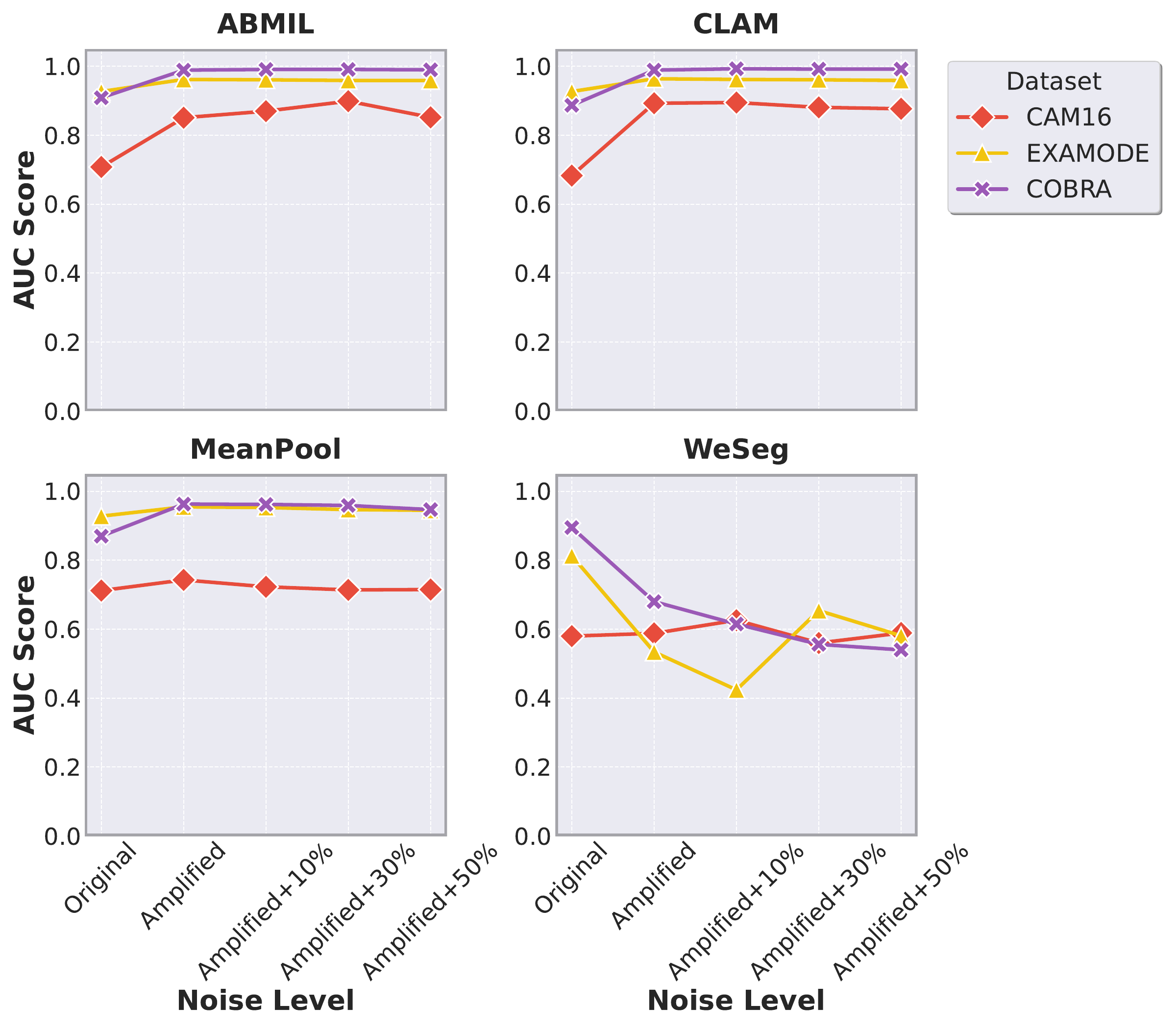} 
  \caption{Impact of amplification on model performance. Line plot showing AUC variation with amplification across noise levels. }
  \label{fig:amp_1}
\end{figure}

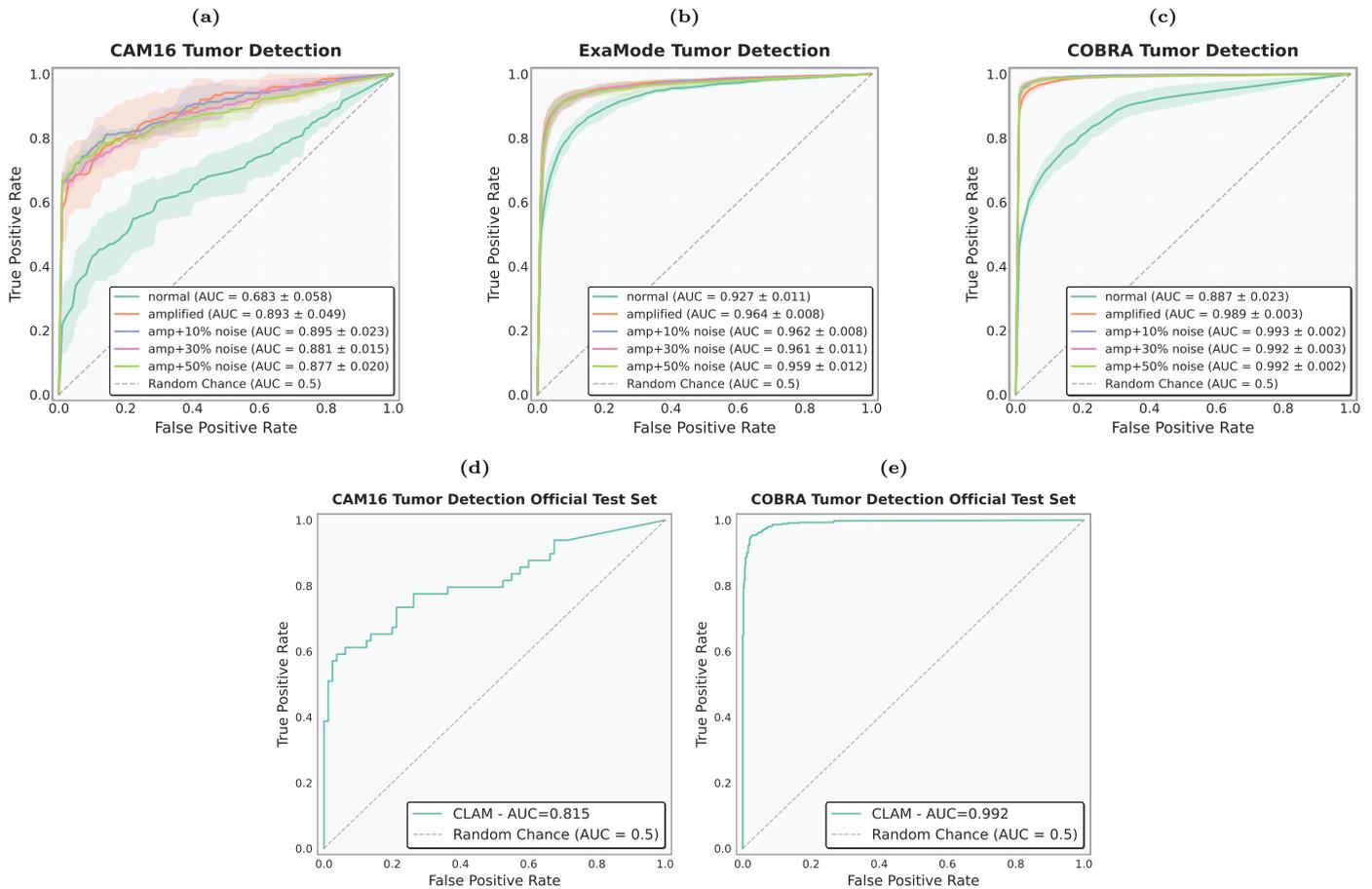
\begin{figure*}[ht]
    \begin{minipage}{1.0\textwidth}
        \centering
        \begin{minipage}{0.3\textwidth}
            \centering
            \begin{tikzpicture}
                \node[anchor=south west, inner sep=0] (image) at (0,0) {\includegraphics[width=\textwidth]{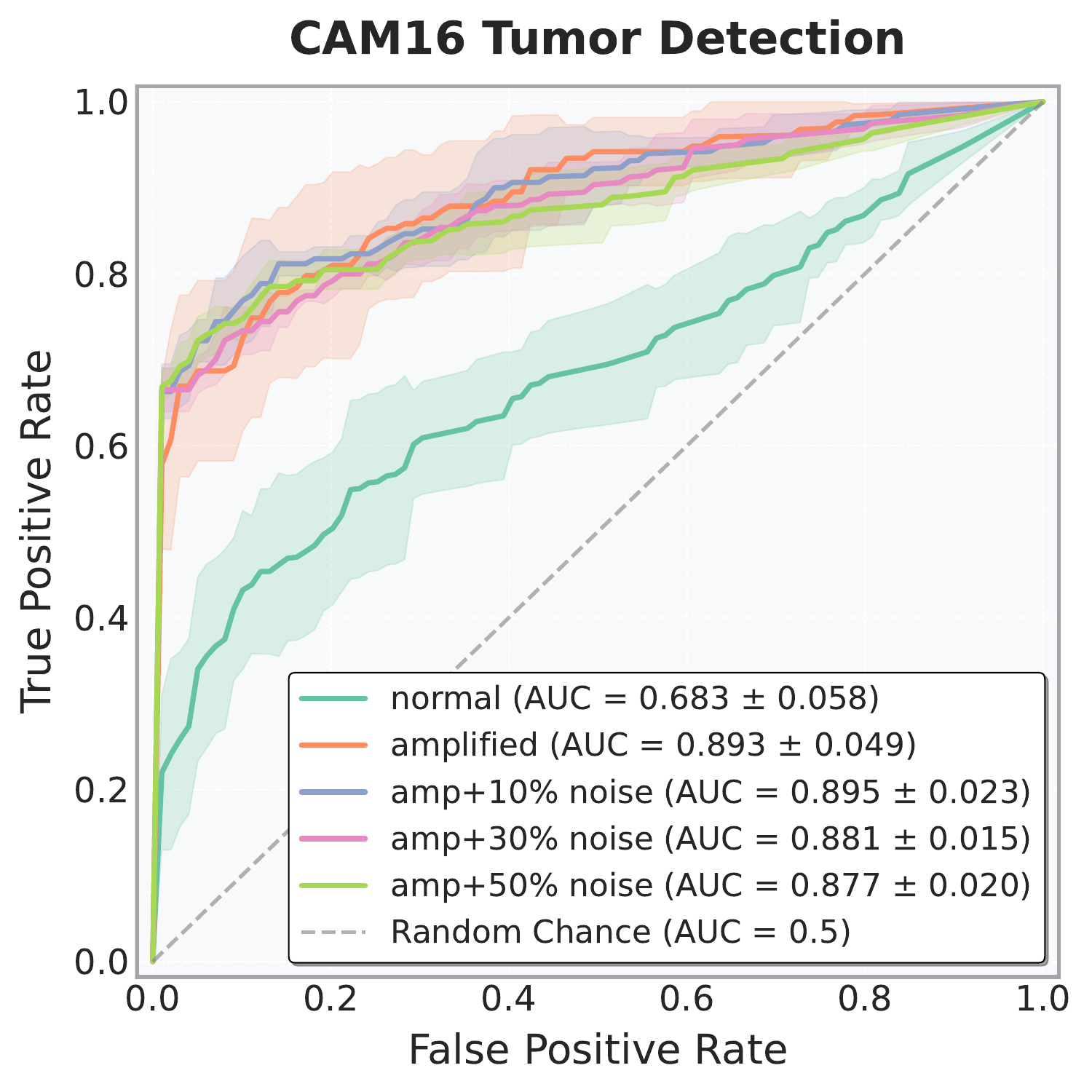}};
                \node[anchor=south, xshift=0.5\textwidth, yshift=0em] at (image.north west) {\scriptsize\textbf{(a)}}; 
            \end{tikzpicture}
            \label{fig:roc1}
        \end{minipage}%
        \hfill
        \begin{minipage}{0.3\textwidth}
            \centering
            \begin{tikzpicture}
                \node[anchor=south west, inner sep=0] (image) at (0,0) {\includegraphics[width=\textwidth]{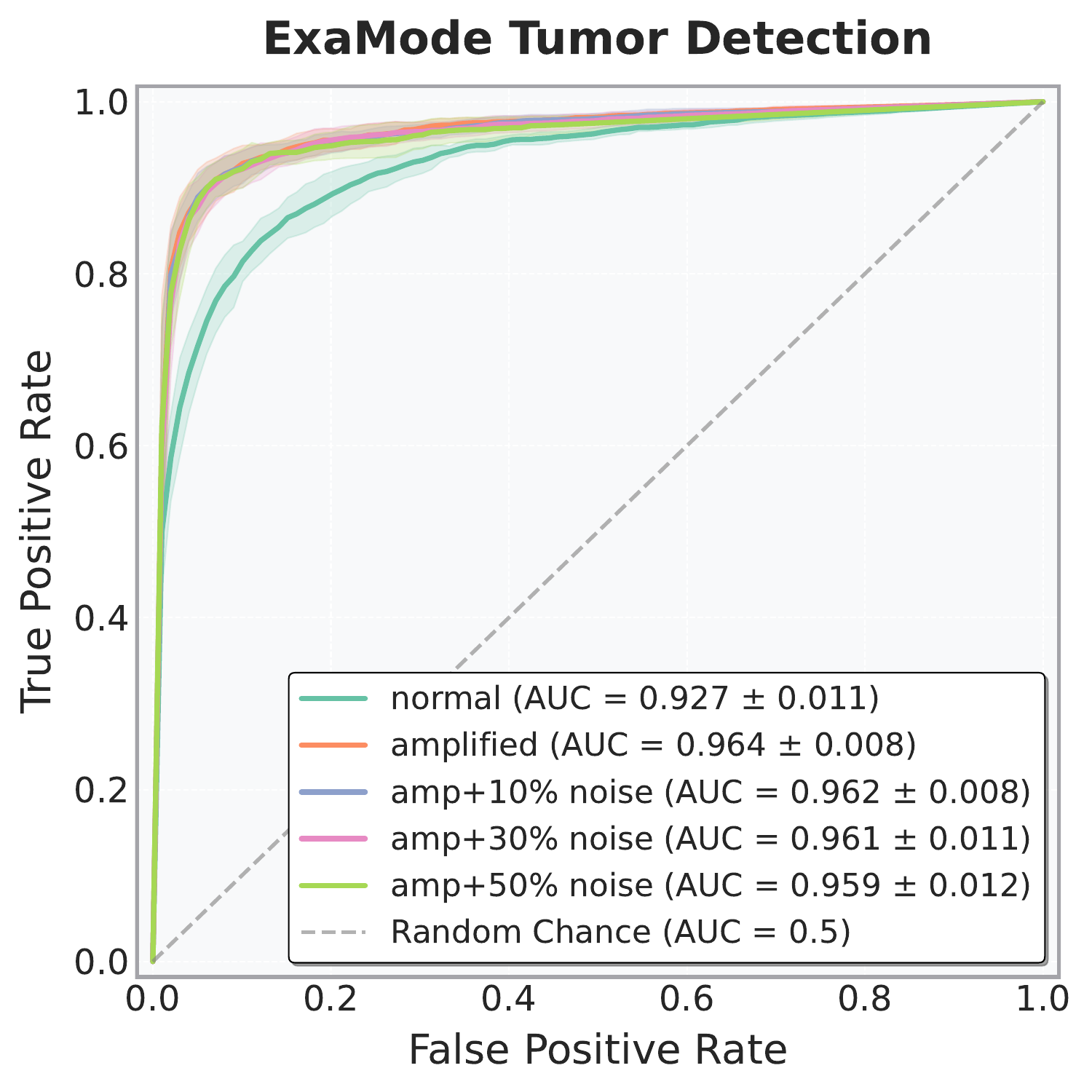}};
                \node[anchor=south, xshift=0.5\textwidth, yshift=0em] at (image.north west) {\scriptsize\textbf{(b)}}; 
            \end{tikzpicture}
            \label{fig:roc2}
        \end{minipage}%
        \hfill
        \begin{minipage}{0.3\textwidth}
            \centering
            \begin{tikzpicture}
                \node[anchor=south west, inner sep=0] (image) at (0,0) {\includegraphics[width=\textwidth]{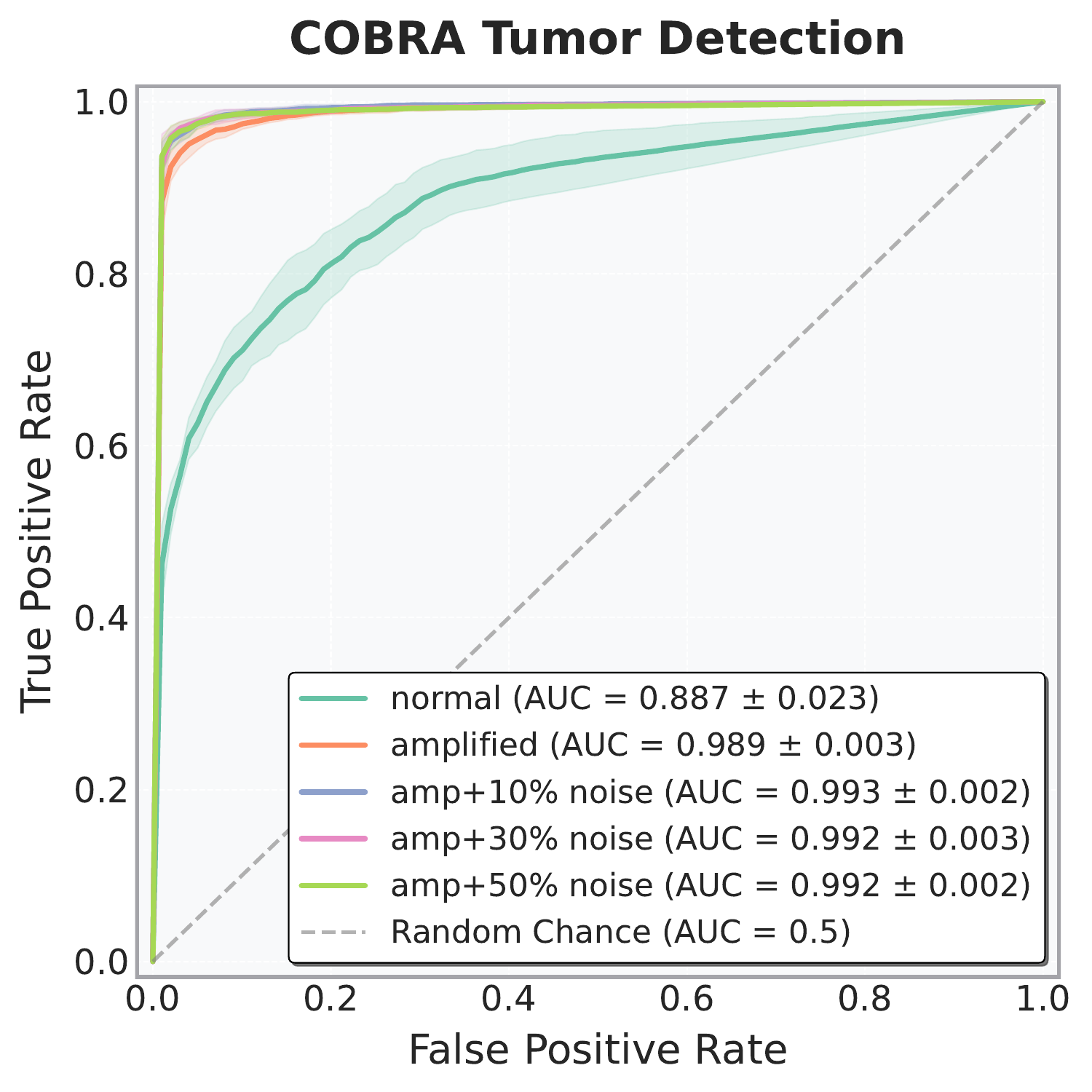}};
                \node[anchor=south, xshift=0.5\textwidth, yshift=0em] at (image.north west) {\scriptsize\textbf{(c)}}; 
            \end{tikzpicture}
            \label{fig:roc3}
        \end{minipage}%
        \vspace{-1em} 
        \begin{minipage}{0.3\textwidth}
            \centering
            \begin{tikzpicture}
                \node[anchor=south west, inner sep=0] (image) at (0,0) {\includegraphics[width=\textwidth]{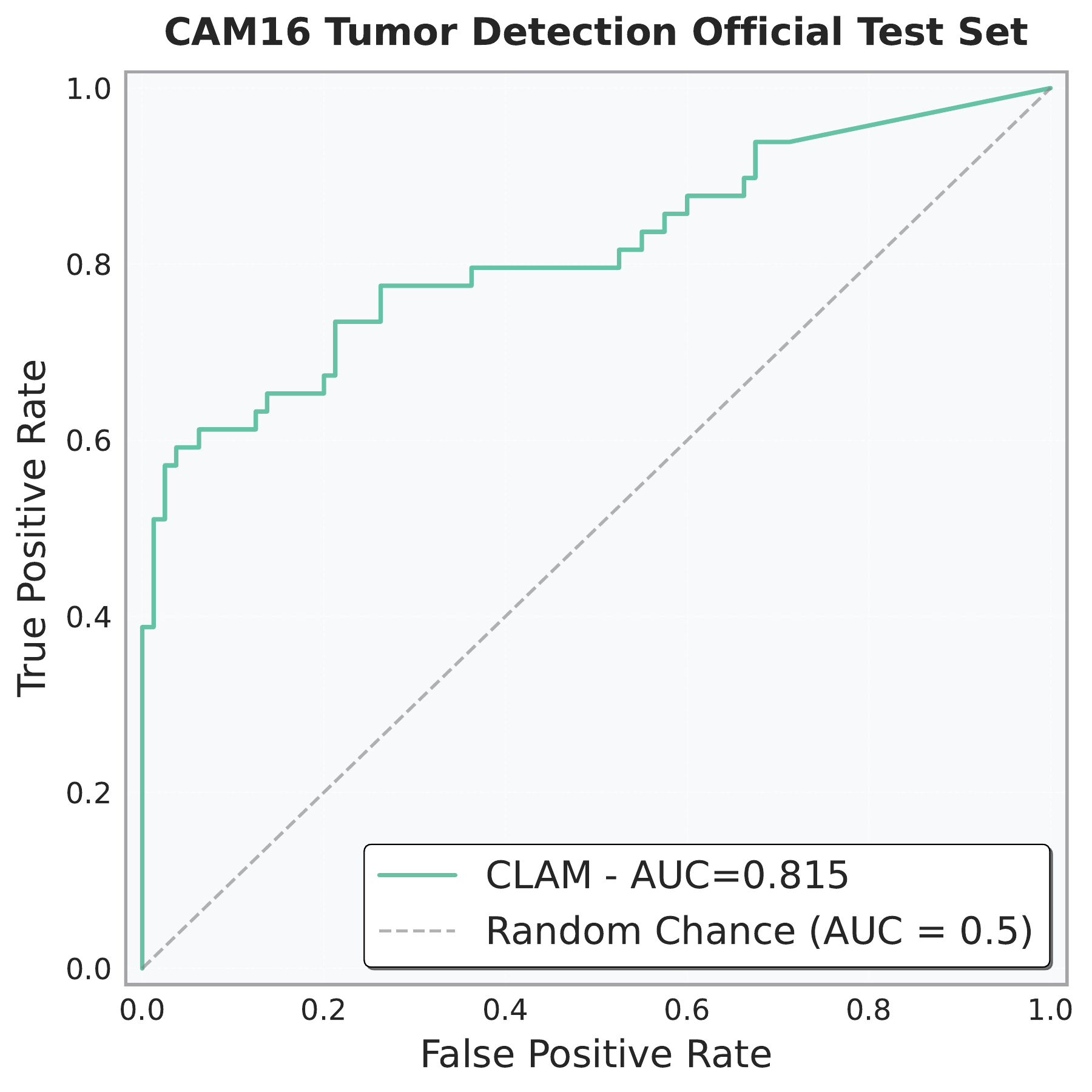}};
                \node[anchor=south, xshift=0.5\textwidth, yshift=0em] at (image.north west) {\scriptsize\textbf{(d)}}; 
            \end{tikzpicture}
            \label{fig:roc4}
        \end{minipage}%
        \hspace{0em} 
        \begin{minipage}{0.3\textwidth}
            \centering
            \begin{tikzpicture}
                \node[anchor=south west, inner sep=0] (image) at (0,0) {\includegraphics[width=\textwidth]{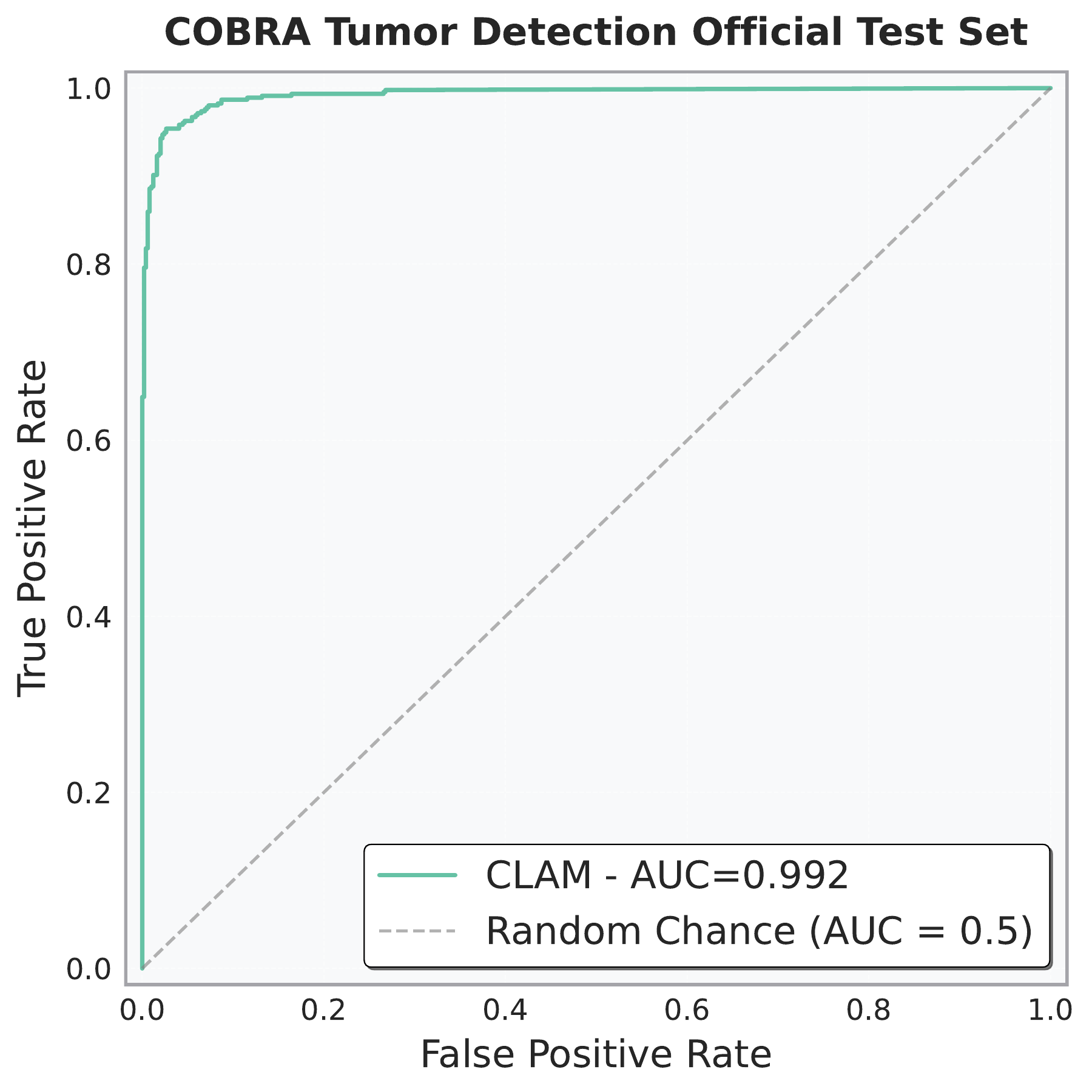}};
                \node[anchor=south, xshift=0.5\textwidth, yshift=0em] at (image.north west) {\scriptsize\textbf{(e)}}; 
            \end{tikzpicture}
            \label{fig:roc5}
        \end{minipage}%
    \end{minipage}

    \caption{(a-c) ROC curves for the CAM16, ExaMode, and COBRA datasets with confidence interval over the 5-fold cross validation, incorporating both amplification and noise effects. (d-e) ROC curves for the CAM16 and COBRA datasets evaluated on the official test sets.}
    \label{fig:combined_plots_amplified}
\end{figure*}

\subsection{Assessing the robustness to noise}
\label{subsec3}
Figure \ref{fig:noise_comparison}(a-b) present the Pearson and Spearman correlation results across different noise levels, showcasing the models' ability to predict tumor percentages under increasing noise. As expected, the models trained with the true tumor percentages exhibit the best overall performance. Despite the expected decrease in performance when noise is added, the results show that the impact of noise is relatively moderate. Notably, performance degradation becomes more pronounced when 50\% noise is introduced, but this represents an extreme and unrealistic scenario in real-world applications. 

For the TNBC dataset, all models demonstrated strong robustness to noise, maintaining high correlation scores even under substantial noise. ABMIL and MeanPool showed particularly robust performance, with Pearson and Spearman correlations decreasing by only 14\% at 50\% noise and maintaining correlations above 0.83. CLAM, despite its strong baseline, showed greater sensitivity, with a 20\% decrease under the same conditions. WeSEG, while starting from a lower baseline correlation, showed robustness with a 12\% decrease. This indicates that its segmentation quality remained effective at identifying tumor areas despite the added noise, indirectly supporting tumor percentage estimation. 

For the AQUILA dataset, models were more affected by noise overall. At 30\% noise, correlations dropped by 17\% for MeanPool (Pearson: 0.964 to 0.804, Sperman: 0.962 to 0.516) and up to 27\% for WeSEG (Pearson: 0.815 to 0.597, Sperman: 0.8 to 0.607). At 50\% noise, ABMIL and MeanPool maintained correlations above 0.5 but experienced 41\% and 46\% declines, respectively. WeSEG, despite its lower baseline correlations, performed comparably to ABMIL and MeanPool under 50\% noise, maintaining correlations around 0.57. Surprisingly, CLAM showed the greatest sensitivity, with correlations declining by approximately 80\%. 

The CAM16 dataset presented the most challenging scenario, with overall lower baseline correlation scores across all models. ABMIL and CLAM emerged as the most robust models, experiencing a 28\% reduction in Pearson correlation at 50\% noise (ABMIL: 0.847 to 0.612, CLAM: 0.811 to 0.596). MeanPool, on the other hand, showed a larger decrease of 37\% (0.876 to 0.552), while WeSEG exhibited the weakest performance with correlations dropping by over 59\%, indicating that the quality of the segmentation is more affected by the challenges posed by small lesions. Spearman correlations showed some unexpected trends, with CLAM's performances remaining stable, while ABMIL and WeSEG showed initial drops at 30\%, followed by an increase of performance at 50\% noise. 

For the ExaMode dataset, models demonstrated strong robustness, with CLAM and MeanPool experiencing a drop of 10\% at 50\% noise (CLAM: 0.95 to 0.849, MeanPool: 0.953 to 0.854), and ABMIL a drop of 13\% (0.95 to 0.827). WeSEG showed higher sensitivity to noise, with a small 6.6\% decline at 30\% noise but a significant 61\% drop at 50\% noise. Interestingly, Spearman correlations for ABMIL and CLAM improved with noise, whereas MeanPool showed slight declines.

Lastly, all models performed robustly on the COBRA dataset, with MeanPool showing only a 6\% drop (0.954 to 0.892) at 50\% noise. CLAM, WeSEG and ABMIL followed with 15\%, 20\% and 21\% decline, respectively. Spearman correlations remained stable or slightly improved for CLAM, ABMIL, and MeanPool, while WeSEG showed minor decreases.

To further assess robustness, we analyzed AUC scores across noise levels for the CAM16, EXAMODE, and COBRA datasets (Figure \ref{fig:noise_comparison}c). ABMIL and CLAM consistently achieved the highest AUC values, underscoring the robustness of attention-based methods. Interestingly, they showed improved results under noise conditions across the three datasets. For ExaMode, ABMIL's AUC increased from 0.927 to 0.947, and CLAM's from 0.927 to 0.943. Similarly, in the COBRA dataset, CLAM's AUC improved from 0.887 to 0.974, and ABMIL's improved from 0.909 to 0.963 at 50\% noise. For the CAM16 dataset, ABMIL showed a modest improvement from 0.708 to 0.732, while CLAM's AUC increased from 0.683 to 0.737. In contrast, WeSEG showed a slight drop of performance on the three datasets, especially on ExaMode (0.813 to 0.608) and COBRA (0.895 to 0.782). 

\begin{table*}[ht]
\centering\
\resizebox{\textwidth}{!}{ 
\begin{tabular}{|l|c|c|c|c|c|c|c|c|c|c|c|}
\hline
\multirow{2}{*}{\textbf{Model}} & \multirow{2}{*}{\textbf{Task}} & \multicolumn{2}{c|}{\textbf{TNBC}} & \multicolumn{2}{c|}{\textbf{AQUILA}} & \multicolumn{2}{c|}{\textbf{CAM16}} & \multicolumn{2}{c|}{\textbf{ExaMode}} & \multicolumn{2}{c|}{\textbf{COBRA}} \\
\cline{3-12}
                               &                               & \textbf{Attention} & \textbf{Logits} & \textbf{Attention} & \textbf{Logits} & \textbf{Attention} & \textbf{Logits} & \textbf{Attention} & \textbf{Logits} & \textbf{Attention} & \textbf{Logits} \\
\hline
\multirow{4}{*}{ABMIL}    & No noise   & 0.539 & 0.939 & 0.568 & 0.98 & 0.511 & 0.915 & 0.648 & 0.863 & 0.738 & 0.786 \\
                          & 10\% noise & 0.383 & 0.927 & 0.457 & 0.967 & 0.719 & 0.951 & 0.591 & 0.853 & 0.737 & 0.781 \\
                          & 30\% noise & 0.325 & 0.898 & 0.219 & 0.887 & 0.865 & 0.959 & 0.625 & 0.822 & 0.727 & 0.693 \\
                          & 50\% noise & 0.542 & 0.879 & 0.353 & 0.793 & 0.763 & 0.886 & 0.715 & 0.8 & 0.732 & 0.664 \\
                          & Amplified & - & - & - & - & 0.841 & 0.945 & 0.738 & 0.786 & 0.849 & 0.899 \\
\hline
\multirow{4}{*}{CLAM}     & No noise   & 0.554 & 0.94 & 0.736 & 0.982 & 0.571 & 0.959 & 0.774 & 0.869 & 0.741 & 0.754 \\
                          & 10\% noise & 0.449 & 0.926 & 0.619 & 0.967 & 0.594 & 0.953 & 0.705 & 0.856 & 0.733 & 0.728 \\
                          & 30\% noise & 0.256 & 0.899 & 0.307 & 0.887 & 0.669 & 0.931 & 0.707 & 0.814 & 0.731 & 0.657 \\
                          & 50\% noise & 0.272 & 0.878 & 0.27 & 0.636 & 0.653 & 0.876 & 0.756 & 0.769 & 0.728 & 0.578 \\
                          & Amplified & - & - & - & - & 0.935 & 0.897 & 0.741 & 0.754 & 0.848 & 0.895 \\
\hline
\multirow{4}{*}{MeanPool} & No noise   &  - & 0.936  & -  & 0.982  & -  & 0.924 &  - & 0.863 & -  & 0.793 \\
                          & 10\% noise & - & 0.925 & - & 0.974 & - & 0.928 & - & 0.851 & -  & 0.787 \\
                          & 30\% noise & - & 0.897 & - & 0.945 & - & 0.933 & - & 0.832 &  - & 0.775 \\
                          & 50\% noise & - & 0.873 & - & 0.789 & - & 0.93 & - & 0.828 & -  & 0.769 \\
                          & Amplified & - & - & - & - & - & 0.924 & - & 0.793 & - & 0.854 \\
\hline
\multirow{4}{*}{WeSEG}    & No noise   & - & 0.83 & -  & 0.877 & - & 0.68 & -  & 0.785 & - & 0.918 \\
                          & 10\% noise & - & 0.885 & - & 0.958 & - & 0.8 &  - & 0.67 & - & 0.89 \\
                          & 30\% noise &  - & 0.842 & - & 0.904 &  - &  0.44 &  - & 0.68 & - & 0.889 \\
                          & 50\% noise & - & 0.846 & -  & 0.835 & - &  0.66 & - & 0.585 & - & 0.871 \\
                          & Amplified & - & - & -  & - & - & 0.74 & - & 0.505 & - & 0.566 \\
\hline
\end{tabular}
}
\caption{This table summarizes the AUC results for attention scores and instance logits compared to the true tumor masks. Results are presented for different methods and datasets for training experiments without noise, with noise, and with amplification.}
\label{tab:metrics}
\end{table*}
\subsection{Assessing the role of the amplification technique}
\label{subsec4}
Figure \ref{fig:amp_1} shows AUC results across datasets and methods under different noise levels, while Figure \ref{fig:combined_plots_amplified}(a-c) highlights ROC curves for the best-performing model, CLAM. As illustrated by the AUC results, the amplification technique significantly improved the performance of attention-based methods such as ABMIL and CLAM. For CAM16, ABMIL's AUC increased from 0.708 to 0.851, and CLAM's from 0.683 to 0.893, demonstrating enhanced detection of small lesions and better differentiation between tumor and tumor-free cases. In contrast, MeanPool and WeSEG showed only minor improvements, with AUC increasing from 0.712 to 0.743 for MeanPool, and from 0.58 to 0.59 for WeSEG. These findings suggest that these models do not benefit from the amplification technique as much as ABMIL and CLAM. We also present the results with the injection of noise, which align with the previous findings where we observed that AUC generally remains stable or even improves with added noise. On CAM16, ABMIL achieved its highest AUC (0.899) at 30\% noise, while CLAM peaked at 0.895 with 10\% noise.

For EXAMODE, the amplification technique similarly showed robust improvements, especially with ABMIL and CLAM where AUC increased to 0.962 and 0.964, respectively. MeanPool showed only minor improvements from 0.928 to 0.955. Interestingly, WeSEG experienced a significant drop in AUC to 0.534, indicating that amplification might not be beneficial for all models. For COBRA, the AUC demonstrated a similar same trend, with attention-based methods consistently outperforming, maintaining AUC always above 0.9, and WeSEG showing a decrease of performance after the amplification.

To further evaluate the benefits of using the amplification technique and provide results comparable to the literature, we conducted an additional experiment on the official test sets for CAM16 and COBRA, in contrast to the previous results computed using the combined test sets from 5-fold cross-validation. We trained two separate models with amplification using the official splits from the publicly available datasets, and the results are presented in Figure \ref{fig:combined_plots_amplified}(d-e). For COBRA, the amplification method achieved an AUC of 0.99 on the test set, aligning closely with the results reported in \cite{Geij24}, where CLAM was applied to binary classification tasks. Similarly, despite CAM16 showing the lowest performance in the original setting, the amplified model still achieved results comparable to those reported in \cite{Doop23} for the corresponding classification task with CLAM, reaching an AUC of 0.81. 

\label{subsec5}
\begin{figure*}[t]
    \centering
    \includegraphics[width=1.0 \linewidth]{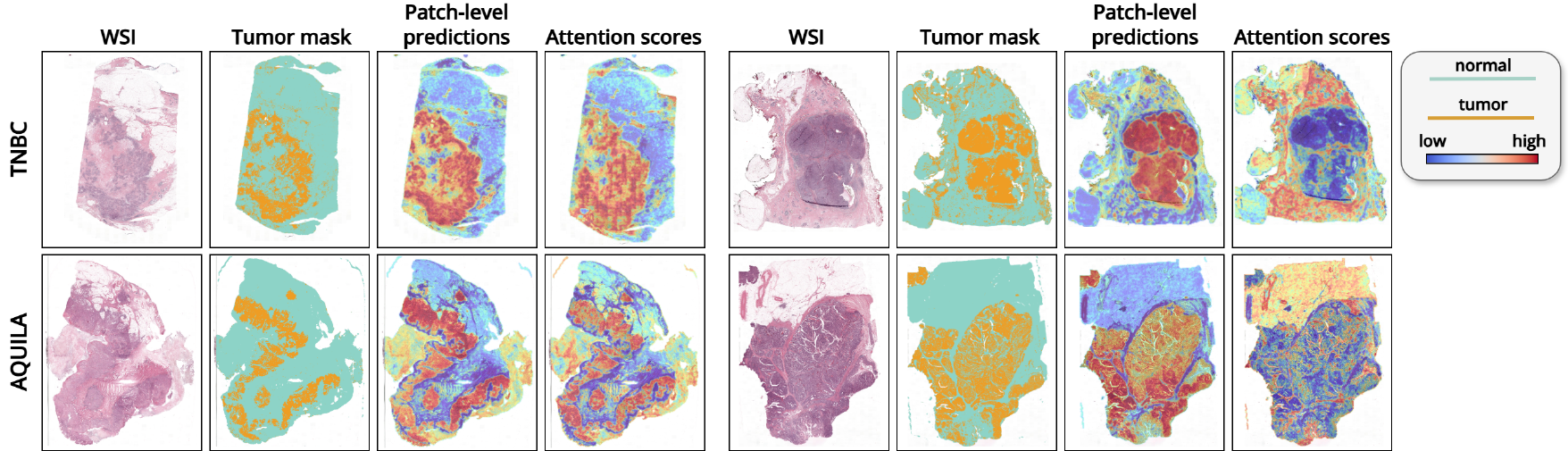} 
    \begin{picture}(0,0)
        \put(-135,170){\small \textbf{(a)}} 
        \put(130,170){\small \textbf{(b)}} 
    \end{picture}
  \caption{Visualization and comparison of logits heatmaps and attention heatmaps for tumor detection. The first row shows examples from the TNBC dataset, while the second row shows examples from the AQUILA dataset. The set of images in column (a) shows good examples where the attention heatmap and logits heatmap align with tumor regions, while column (b) shows examples with reversed attention heatmaps, where the attention mechanism highlights non-tumor regions instead.}
  \label{fig:heatmaps_1}
\end{figure*}
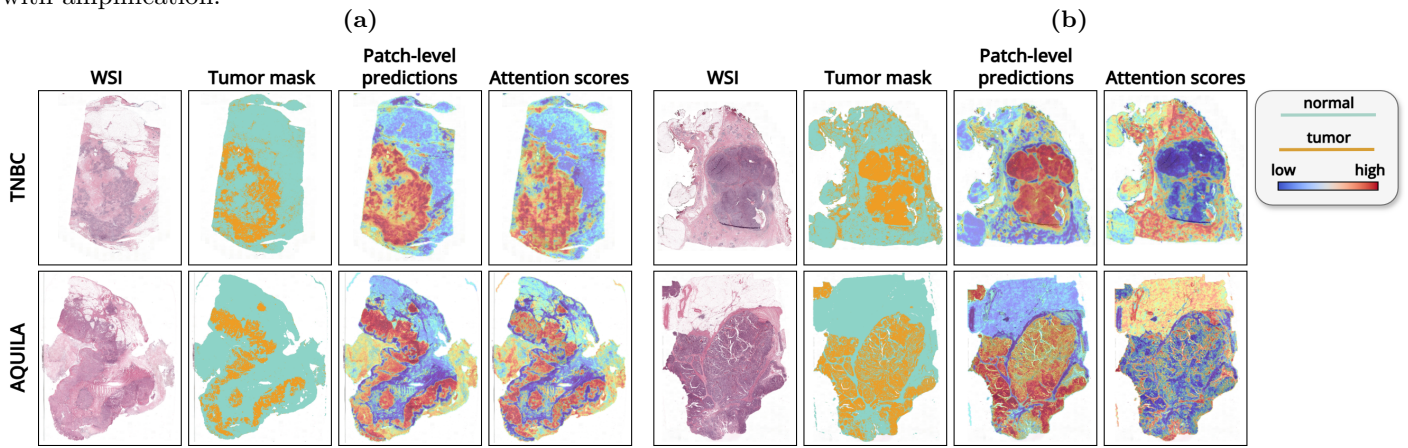

\begin{figure*}[t]
    \centering
    \includegraphics[width=1.0 \linewidth]{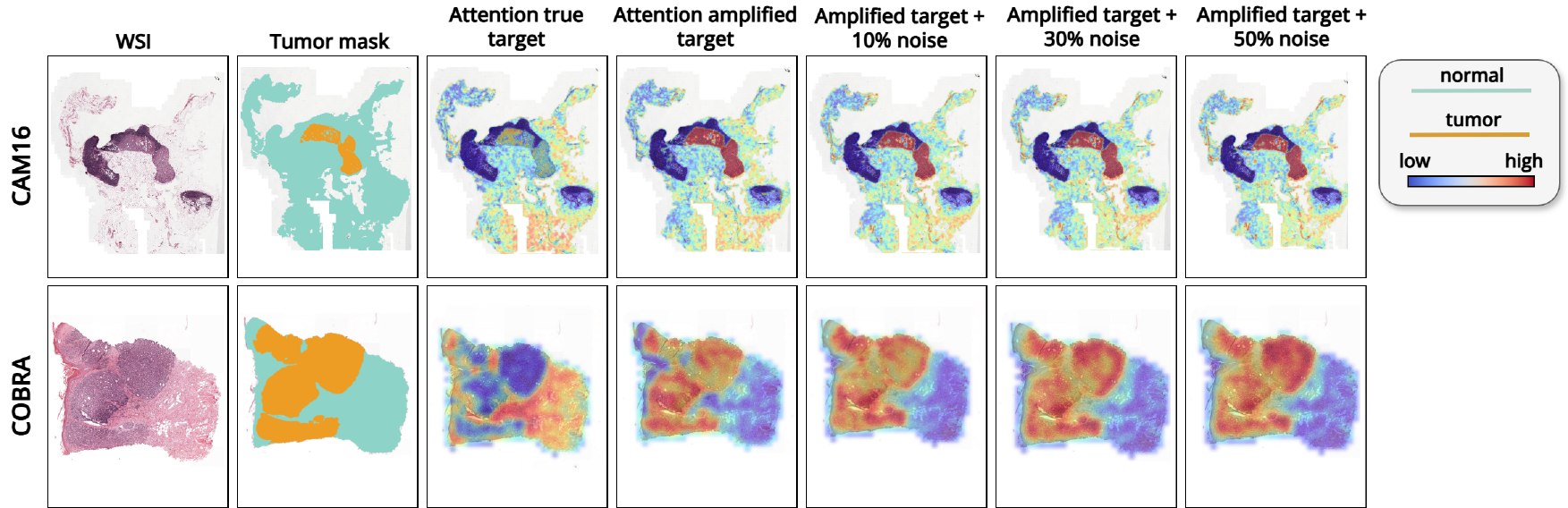} 
  \caption{Visualization and comparison of attention heatmaps for tumor detection, illustrating the effects of the amplification technique and noise on the qualitative results. The figure demonstrates how amplification and increasing noise levels affect the attention mechanism’s ability to highlight tumor regions.}
  \label{fig:heatmaps_2}
\end{figure*}
\subsection{Visual interpretability}
To evaluate interpretability, we first extract patch-level labels by assigning binary labels (0 or 1) to image patches based on the percentage of tumor area within each patch. A threshold of 50\% tumor content is used, where patches with more than 50\% tumor area are labeled as tumor. These patch-level labels serve as the reference standard, based on the tumor annotations or segmentation masks, to quantify the tumor presence within each patch. We then compare these patch-level labels with the instance logits and attention scores, and compute the AUC for each method.

The results in Table \ref{tab:metrics} reveal a significant performance gap. Interestingly, instance logits consistently yield much higher AUC scores across various noise levels and datasets compared to attention scores. Attention scores show considerable variability, with values ranging from low to moderate. To further investigate this phenomenon, we analyzed the attention heatmaps and identified an interesting behavior. In some cases, both the attention heatmaps and instance logits heatmaps closely align with the segmentation map, effectively focusing on tumor regions, as shown in Figure \ref{fig:heatmaps_1}a. However, in other instances, as shown in Figure~\ref{fig:heatmaps_1}b, the instance logits correctly focus on the tumor areas, while the attention heatmap mistakenly assigns higher scores to normal tissue and lower scores to the tumor, producing an "inverted attention heatmap". These variations lead to lower overall performance for attention scores and highlight their limitations in regression tasks. While attention mechanisms have proven highly effective in classification tasks for similar tumor detection scenarios, our findings suggest significant limitations when applied to regression problems. 

Additionally, we evaluate the impact of amplification and noise on interpretability, as shown in Figure \ref{fig:heatmaps_2}. Amplified models produce more consistent and better-aligned heatmaps, with attention focused more accurately on tumor regions. Additionally, the issue of "inverted attention heatmaps" observed in some cases, appears to be mitigated with amplification. Importantly, the qualitative results remain consistent across varying noise levels, with heatmaps showing no degradation even at higher noise levels. 

\label{subsec6}
\section{Discussion and conclusion}
In this study, we introduced a novel weakly-supervised regression framework for tumor detection in whole-slide images, leveraging tumor percentage as a clinically relevant target. While precise annotations or AI-generated segmentation masks were used in this study to compute ground truth tumor percentages, these percentages can be easily estimated from the clinically available coarse annotations of the tumor burden routinely documented in molecular pathology workflows. By shifting the focus from traditional binary classification to regression, this framework addresses the challenges posed by datasets with scarce or absent negative cases, offering a flexible alternative for tumor detection.

Our experiments demonstrated that weakly-supervised models can effectively learn to estimate tumor percentages across diverse datasets. More importantly, we showed that these predicted percentages can serve as a proxy for tumor detection, the primary aim of this framework. However, we recognize that clinical annotations of tumor burden inherently contain noise, introducing variability into the reference standard. To address this, we simulated real-world noise in clinical practice by introducing synthetic noise into our data. Our results showed that the regression framework remained robust even under significant noise levels, especially for attention-based methods (ABMIL and CLAM), demonstrating its potential usability despite noisy annotations. However, it is worth noting that our noise simulations, conducted using uniform distributions with three noise levels ($\pm10\%, \pm30\%, \pm50$\%), may not fully reflect the exact distribution of variability found in clinically available annotations. Future works could aim at better modeling the noise encountered in clinical practice and integrating real-world noisy annotations to better understand the model’s robustness.

Addressing tumor detection in datasets with small lesions proved to be challenging due to the difficulty of the regression framework in distinguishing between small percentages and tumor-free cases, resulting in suboptimal performance compared to classification benchmarks. To mitigate this, we introduced a target amplification technique, which proved particularly valuable for datasets containing biopsies or subtle lesions like CAM16, ExaMode, and COBRA. While target amplification improves the model's sensitivity to small tumor percentages, it introduces a trade-off by compressing higher-end values, which may reduce the model’s precision in predicting tumor percentages for cases with extensive lesions. This limitation could affect the accuracy of regression predictions for large tumors. However, it is important to emphasize that the ultimate goal of this framework is robust tumor detection, not precise tumor percentage regression. The regression task is primarily used as a tool to enable training without relying on negative cases, but the end goal remains the accurate identification of tumor regions.

An important focus of this study was on model interpretability. Attention mechanisms, while widely used in MIL models for tumor segmentation in classification tasks, have been less explored in regression tasks. Interestingly, in our regression framework, attention scores exhibited a counterintuitive behavior, often assigning lower scores to tumor regions and higher scores to normal regions, resulting in "inverted attention heatmaps". Despite this, instance-based MIL frameworks allowed us to compare attention scores with patch-level predictions, which proved to be more effective in localizing tumor regions. Nevertheless, the interpretability of attention scores for regression models remains an open challenge. While the amplification technique improved the performance of attention heatmaps, potentially mitigating the reversed behavior observed in unamplified settings, attention mechanisms may require further refinement and adaptation for use in regression tasks, where the task is to predict a continuous value rather than making binary classifications. 

Our approach has limitations. For this work, we relied on ImageNet pre-trained encoders. Although ImageNet pretraining is a standard practice and has shown general utility in histopathology task, using domain-specific pretraining could provide a better could provide a better representation of histopathology data. Future work could explore leveraging encoders pretrained on large-scale H\&E-stained histopathology datasets, which may better capture domain-specific features and improve generalizability. 
Another area for improvement lies in the interpretability of attention-based heatmaps. Future research should focus on designing more robust interpretability techniques tailored for regression, which could provide clearer insights into the model’s decision-making process. 
Additionally, the fifth-root transformation used for target amplification was empirically effective but may not be the optimal choice for all datasets. Further investigations could explore alternative amplification strategies, or even dynamic amplification methods that adapt based on the specific characteristics of the dataset. Finally, while we simulated noise to evaluate robustness, a reader study with multiple pathologists annotating the tumor burden on the same slides would help quantify the actual variability in clinical annotations and provide valuable insights into the practical applicability of this framework.

\section*{Acknowledgments}
We would like to thank Daan Geijs for the development of the skin segmentation model, John-Melle Bokhorst for the development of the colon segmentation model and Mart Rijthoven for the development of the breast segmentation model. This project has received funding from the Innovative Medicines Initiative 2 Joint Undertaking under grant agreement No 945358. This Joint Undertaking receives support from the European Union’s Horizon 2020 research and innovation program and EFPIA (\url{www.imi.europe.eu}).

\section{Declaration of Generative AI and AI-assisted technologies in the writing process}
During the preparation of this work the author(s) used ChatGPT in order to improve readability and language. After using this tool/service, the author(s) reviewed and edited the content as needed and take(s) full responsibility for the content of the publication.

\bibliographystyle{model2-names.bst}
\bibliography{refs}

\end{document}